%% file: main.tex
\documentclass[aps,pra,reprint,superscriptaddress,floatfix,longbibliography,amsmath,amssymb,raggedbottom]{revtex4-2}

\usepackage{graphicx}
\usepackage{mathrsfs}
\usepackage[hidelinks]{hyperref}
\hypersetup{colorlinks=true,linkcolor=blue,citecolor=blue,urlcolor=blue}

\begin{document}

\title{Coherent state transport for nonreversible Markov chains}

\author{Wael Drif}
\email{wael.drif@keio.jp}
\affiliation{Graduate School of Science and Technology, Keio University,\\
3-14-1 Hiyoshi, Kohoku, Yokohama, Kanagawa 223-8522, Japan}

\begin{abstract}
Quantum algorithms often need a coherent encoding of the stationary distribution of a Markov chain. For nonreversible chains, the available spectral operations may instead select a different state. We give an error-controlled multistage framework for transporting these states, which are Perron states in our forward-only construction. Adjacent selected-state overlaps govern transport, while only the initial and final biases enter additively into the final stationary-state error bound. We certify overlaps using stationary-state geometry or changes in flat discriminants, the latter requiring no intermediate stationary-state information. An exact filter-response identity controls selective-phase errors without assuming commutation with the target projector. A bounded Laurent filter implements the forward-only phases with one signal qubit, with degree governed by the Poincar\'e gap under suitable spectral conditions. In an eight-state nonreversible Gibbs family, we bound the preparation error below 0.001 even when an intermediate bias exceeds this tolerance, under stated implementation tolerances. We also certify spectral targets and overlaps uniformly over a temperature interval.
\end{abstract}

\maketitle

\section{Introduction}
\label{sec:introduction}

Quantum algorithms that operate on probability distributions often require their coherent amplitude encodings rather than individual classical samples. For a stationary distribution $\pi$ on a finite state space $\Omega$, the corresponding quantum sample (QSAMPLE) is
\begin{equation}
|\pi\rangle=\sum_{x\in\Omega}\sqrt{\pi(x)}|x\rangle.
\end{equation}

Measuring this state produces a sample from $\pi$, while retaining coherence allows its use in amplitude estimation, quantum Monte Carlo, and simulated annealing \cite{Somma2008,WocjanAbeyesinghe2008,Montanaro2015,Arunachalam2022,Brassard2002}. In quantum simulated annealing, preparation proceeds through a sequence of such states as the target distribution changes \cite{Somma2008,WocjanAbeyesinghe2008,WocjanChiangNagajAbeyesinghe2009,HarrowWei2020}. Nonreversible Markov chains are of particular interest in this setting, as their stationary probability currents can alter relaxation behavior and accelerate mixing in suitable models \cite{ChenLovaszPak1999,DiaconisHolmesNeal2000,EberleLorler2026,Gao2020}.

The spectral structure of a Markov chain determines which coherent states can be selected by quantum-walk operations. For reversible chains, the stationary QSAMPLE can be isolated in the relevant walk sector, enabling state-selective reflections and their use in quantum sampling \cite{MagniezNayakRolandSantha2011,WocjanAbeyesinghe2008,Szegedy2004}. For nonreversible chains, Claudon, Piquemal, and Monmarch\'e \cite{Claudon2025} established two complementary spectral constructions: a curved discriminant whose singular-value-one target is the stationary QSAMPLE, accessed using coherent forward and time-reversed transitions, and a symmetric flat discriminant whose leading Perron state can be selected using forward transitions alone. Separately, Zhao \cite{Zhao2026} developed an ancilla-efficient multistage preparation method for reversible chains, combining generalized quantum signal processing (GQSP) \cite{MotlaghWiebe2024} with fixed-point amplitude amplification (FPAA) \cite{YoderLowChuang2014}.

Multistage preparation between arbitrary overlapping states was established by Wocjan and Abeyesinghe \cite{WocjanAbeyesinghe2008}. For nonreversible chains, however, the selected states need not be the stationary QSAMPLEs. Stationary-state overlaps alone therefore do not certify the cost of transport.

We give a common framework for certifying and implementing transport between the states actually selected. In the forward-only route, these are Perron states of flat discriminants. Only the endpoint biases enter additively into the sufficient final stationary-state error bound. We certify adjacent Perron overlaps from stationary-state geometry or directly from changes in discriminants. The direct certificate requires neither intermediate stationary-state overlaps nor intermediate bias bounds.

An exact response identity controls the selective-phase error without requiring commutation with the target projector. The forward-only construction selects the complete conjugate Perron eigenphase pair through a directly normalized bounded Laurent filter using one GQSP signal qubit. Its filtering guarantee depends on the Poincar\'e gap. An eight-state Gibbs family illustrates certified transport through a biased intermediate Perron state without using intermediate stationary-state information in the transport certificates.

\section{Coherent transport and stationary-state accuracy}
\label{sec:transport}

\subsection{Coherent access and selected states}

Consider a sequence of ergodic Markov kernels $P_0,\ldots,P_\ell$ on a finite state space $\Omega$, with full-support stationary distributions $\pi_0,\ldots,\pi_\ell$. At each stage, coherent access to a transition kernel $P_i$ is provided by a unitary $O_{P_i}$ satisfying
\begin{equation}
O_{P_i}|x\rangle|0^n\rangle
=
|x\rangle\sum_{y\in\Omega}
\sqrt{P_i(x,y)}\,|y\rangle,
\label{eq:coherent-transition-oracle}
\end{equation}
where $n=\lceil\log_2|\Omega|\rceil$, together with its inverse and the controlled versions used by the spectral circuits. Powered and time-reversed kernels require separately supplied or constructed clean endpoint oracles; neither follows generally from $O_{P_i}$. The desired stationary sample is $|\pi_i\rangle$, with projector $\Pi_{\pi_i}=|\pi_i\rangle\langle\pi_i|$, and preparation accuracy is measured by trace distance $D_{\mathrm{tr}}(\rho,\Pi_{\pi_i})=\frac12\|\rho-\Pi_{\pi_i}\|_1$; this also bounds the total-variation error after computational-basis measurement. The coherent-access conventions and complete discriminant encodings are developed in Appendix~\ref{app:access}.

The state isolated by a spectral construction depends on the available transition access. The curved discriminant selects the stationary state $|\pi_i\rangle$, whereas the forward-only flat discriminant selects its normalized Perron state $|\mu_i\rangle$. Let
\begin{equation}
|\sigma_i\rangle=
\begin{cases}
|\pi_i\rangle, & \text{curved route}\\
|\mu_i\rangle, & \text{flat route}
\end{cases}
\label{eq:selected-state}
\end{equation}
and $\Pi_i=|\sigma_i\rangle\langle\sigma_i|$. The ideal selective phase is
\begin{equation}
S_\phi(\Pi_i)
=
I+(e^{i\phi}-1)\Pi_i.
\label{eq:ideal-selective-phase}
\end{equation}

The quantity governing transport between consecutive stages is
\begin{equation}
q_i
:=
|\langle\sigma_i|\sigma_{i+1}\rangle|^2.
\label{eq:selected-overlap}
\end{equation}

\begin{figure}[t]
\includegraphics[width=\columnwidth]{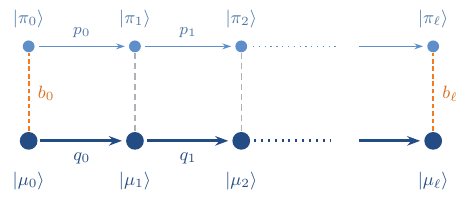}
\caption{Coherent transport of selected states. The upper path contains the stationary QSAMPLEs $|\pi_i\rangle$ with adjacent squared overlaps $p_i$; the lower path contains the selected Perron states $|\mu_i\rangle$ with squared overlaps $q_i$. Orange dashed connectors mark the endpoint biases $b_0$ and $b_\ell$, which enter the sufficient final stationary-state error bound. Gray dashed connectors indicate interior biases, which affect the transport overlaps. Dotted segments represent omitted stages. The diagram is schematic.}
\label{fig:transport}
\end{figure}

\subsection{Multistage coherent transport}

To compose the selective phases across different chains, let $\mathcal H_{\mathrm{sys}}$ be the state space and $\mathcal H_{\mathrm{work}}$ include the system and all auxiliary registers. We use a fixed clean-register isometry
\begin{equation*}
J:\mathcal H_{\mathrm{sys}}\longrightarrow
\mathcal H_{\mathrm{work}},
\qquad J^\dagger J=I,
\end{equation*}
shared by all stages, which embeds the system with the same zero-initialized work registers at every stage. For a compiled selective phase $\widetilde S_{\phi,i}$, require
\begin{equation}
\left\|
\widetilde S_{\phi,i}J
-
JS_\phi(\Pi_i)
\right\|
\leq \eta_{\phi,i}.
\label{eq:phase-error-bound}
\end{equation}

This condition provides a common interface for both spectral routes.

For each transition $i\to i+1$, suppose a certified overlap lower bound
\begin{equation*}
0<\underline q_i\leq q_i
\end{equation*}
is available. We measure primitive-oracle cost in calls to the specified coherent transition-oracle primitives; a clean powered-transition oracle is charged according to its assigned implementation cost, such as $c_{P^m}$. Optimal FPAA \cite{YoderLowChuang2014} uses an odd FPAA parameter
\begin{equation}
L_i
=
O\!\left(
\frac{1}{\sqrt{\underline q_i}}
\log\frac{1}{\epsilon_i^{\mathrm{FP}}}
\right)
\label{eq:fpaa-length}
\end{equation}
to transport $|\sigma_i\rangle$ toward $|\sigma_{i+1}\rangle$, with intrinsic trace-distance error at most $\epsilon_i^{\mathrm{FP}}$. Let $\eta_i^{\mathrm{ph}}$ bound the compiled phase errors at both endpoints of that transition.

\textbf{Theorem 1 (Coherent transport of selected states).}
Let $|\sigma_0\rangle,\ldots,|\sigma_\ell\rangle$ be normalized states with adjacent squared overlaps $q_i>0$ and certified bounds $0<\underline q_i\leq q_i$. Suppose the compiled selective phases satisfy the common clean-register condition above at every stage. For an initial state $\rho_0$, define
\begin{equation*}
\delta_0
:=
D_{\mathrm{tr}}(\rho_0,\Pi_0).
\end{equation*}

After the sequence of FPAA transformations, the final state $\rho_\ell$ satisfies
\begin{equation}
D_{\mathrm{tr}}(\rho_\ell,\Pi_\ell)
\leq
\delta_0+
\sum_{i=0}^{\ell-1}
\left[
2(L_i-1)\eta_i^{\mathrm{ph}}
+
\epsilon_i^{\mathrm{FP}}
\right].
\label{eq:transport-error}
\end{equation}

If $C_i$ denotes the primitive-oracle cost of one implemented selective phase about $|\sigma_i\rangle$, the total primitive-oracle cost is
\begin{equation}
\widetilde O\!\left(
\sum_{i=0}^{\ell-1}
\frac{\max\{C_i,C_{i+1}\}}
{\sqrt{\underline q_i}}
\right),
\label{eq:transport-complexity}
\end{equation}
where $\widetilde O$ suppresses the logarithmic accuracy factors in the filter and FPAA implementations.

Theorem 1 applies the arbitrary-state transport of Ref.~\cite{WocjanAbeyesinghe2008} using optimal FPAA \cite{YoderLowChuang2014} and the common embedding of Ref.~\cite{Zhao2026}, with explicit errors for imperfect spectral selectors. Clean-register phase errors telescope along the ideal trajectory. The bound uses the conservative factor $2(L_i-1)$; the standard sequence contains $(L_i-1)/2$ phases about each endpoint. The proof in Appendix~\ref{app:transport} applies to any sequence of selectors satisfying the common clean-register condition.

\subsection{Endpoint bias and overlap geometry}

For the flat route, the bias between the selected Perron state and the stationary QSAMPLE is
\begin{equation}
b_i
:=
D_{\mathrm{tr}}
\left(
|\mu_i\rangle\langle\mu_i|,
\Pi_{\pi_i}
\right)
=
\sqrt{1-|\langle\mu_i|\pi_i\rangle|^2}.
\label{eq:perron-bias}
\end{equation}

Starting from $|\pi_0\rangle$ gives initial mismatch $b_0$, and comparison of the transported final Perron state with $|\pi_\ell\rangle$ contributes $b_\ell$. Theorem 1 and the triangle inequality give
\begin{equation}
D_{\mathrm{tr}}(\rho_\ell,\Pi_{\pi_\ell})
\leq
b_0+b_\ell+
\sum_{i=0}^{\ell-1}
\left[
2(L_i-1)\eta_i^{\mathrm{ph}}
+
\epsilon_i^{\mathrm{FP}}
\right].
\label{eq:endpoint-error}
\end{equation}

As illustrated in Fig.~\ref{fig:transport}, intermediate biases affect the path through the overlaps of the selected states, whereas only the initial and final biases enter additively in this sufficient stationary-state accuracy bound.

To connect a schedule designed for stationary distributions to the forward-only transport path, define
\begin{equation}
p_i
:=
|\langle\pi_i|\pi_{i+1}\rangle|^2,
\qquad
q_i
=
|\langle\mu_i|\mu_{i+1}\rangle|^2.
\label{eq:stationary-selected-overlaps}
\end{equation}

For normalized pure states, the Fubini--Study distance is $d_{\mathrm{FS}}(\psi,\varphi)=\arccos|\langle\psi|\varphi\rangle|$. Since $d_{\mathrm{FS}}(\pi_i,\mu_i)=\arcsin b_i$, its triangle inequality gives
\begin{equation}
\begin{aligned}
q_i&\geq\cos^2\!\left(\min\{\pi/2,\theta_i\}\right),\\
\theta_i&=\arccos\sqrt{p_i}+\arcsin b_i+\arcsin b_{i+1}.
\end{aligned}
\label{eq:overlap-bound}
\end{equation}

This bound is sharp over normalized pure states given only a stationary-overlap lower bound and bias upper bounds. Appendix~\ref{app:transport} gives the proof; Sec.~\ref{sec:resources} and Appendix~\ref{app:certificates} give the spectral overlap certificates.

\section{Spectral realization of selective phases}
\label{sec:spectral}

\subsection{Transition encodings and the selective-phase interface}
\label{subsec:phase-interface}

A coherent transition oracle defines the isometry
\begin{equation*}
\square_P|x\rangle
=
|x\rangle\sum_y\sqrt{P(x,y)}\,|y\rangle.
\end{equation*}

With the register swap $S|x\rangle|y\rangle=|y\rangle|x\rangle$, the compression
\begin{equation*}
D(P_1,P_2)
:=
\square_{P_1}^{\dagger}S\square_{P_2}
\end{equation*}
has matrix elements
\begin{equation*}
D(P_1,P_2)(x,y)
=
\sqrt{P_1(x,y)P_2(y,x)}.
\end{equation*}

Thus transition-state preparation, swap, and uncomputation provide the projected unitary encoding used by both routes \cite{Claudon2025,Szegedy2004}.

Let a filtering unitary $C$ have accepted block $A$, namely the block for which the signal qubit enters and exits in $|0\rangle$, and response $G=A^\dagger A$. Define
\begin{equation*}
P_\phi
=
e^{i\phi}|0\rangle\langle0|
+
|1\rangle\langle1|,
\qquad
U_\phi
=
C^\dagger(P_\phi\otimes I)C.
\end{equation*}

Thus $U_\phi$ applies a phase to the successful filter outcome and uncomputes the filter, as in Ref.~\cite{Zhao2026}, producing an approximate selective phase about a target projector $\Lambda$.

\textbf{Proposition 1 (Selective-phase response).}
Let $\Lambda$ be an orthogonal projector and $\Lambda^\perp=I-\Lambda$. Define the target rejection and bulk acceptance errors by
\begin{equation*}
\begin{aligned}
\delta_{\mathrm{tar}}&=\|\Lambda(I-G)\Lambda\|,\\
\delta_{\mathrm{bulk}}&=\|\Lambda^\perp G\Lambda^\perp\|.
\end{aligned}
\end{equation*}

Then
\begin{equation}
\begin{aligned}
&\left\|
U_\phi(|0\rangle\otimes I)
-
|0\rangle\otimes S_\phi(\Lambda)
\right\|\\
&\qquad=
|e^{i\phi}-1|
\sqrt{\max\{\delta_{\mathrm{tar}},\delta_{\mathrm{bulk}}\}}.
\end{aligned}
\label{eq:selective-phase-response}
\end{equation}

No commutation between $G$ and $\Lambda$ is required. If $\Lambda G\Lambda\geq q_{\mathrm{tar}}\Lambda$ and $\Lambda^\perp G\Lambda^\perp\leq q_{\mathrm{bulk}}\Lambda^\perp$, the error is bounded by $|e^{i\phi}-1|\sqrt{\max\{1-q_{\mathrm{tar}},q_{\mathrm{bulk}}\}}$. Appendix~\ref{app:phase-response} illustrates the filter--phase--uncompute circuit and proves the identity for arbitrary input isometries and orthogonal success projectors.

\subsection{Forward-only selection by a bounded Laurent filter}
\label{subsec:flat-route}

For a positive integer power $j$, the flat construction uses
\begin{equation*}
D_j
:=
D(P^j,P^j),
\qquad
D_j(x,y)
=
\sqrt{P^j(x,y)P^j(y,x)}.
\end{equation*}

Suppose $D_j$ is primitive. Its normalized positive Perron vector satisfies
\begin{equation}
D_j|\mu_j\rangle
=
\omega_j|\mu_j\rangle,
\qquad 0<\omega_j\leq1.
\label{eq:flat-perron-eigenpair}
\end{equation}

Writing $\mu_j(x)=\langle x|\mu_j\rangle>0$, the geometric reversibilization is the reversible Perron similarity transform of $D_j/\omega_j$,
\begin{equation*}
Q_j(x,y)
=
\frac{D_j(x,y)\mu_j(y)}
{\omega_j\mu_j(x)},
\end{equation*}

$Q_j$ is stochastic and reversible with stationary distribution $\nu_j(x)=\mu_j(x)^2$, the most-reversible distribution of Ref.~\cite{Claudon2025}. Its Poincar\'e gap is
\begin{equation}
\gamma_+(Q_j)
=
1-\frac{\lambda_2(D_j)}{\omega_j}.
\label{eq:flat-spectral-gap}
\end{equation}

Here $\lambda_2(D_j)$ denotes the second-largest eigenvalue in algebraic order. Using the Poincar\'e gap rather than the absolute gap, we obtain the following variant of the leading-eigenvalue selection condition of Ref.~\cite{Claudon2025}:
\begin{equation}
\omega_j
\geq
\frac{1-c_\star}
{1-c_\star[1-\gamma_+(Q_j)]},
\qquad c_\star=\frac1{16}.
\label{eq:leading-eigenvalue-condition}
\end{equation}

Indeed, with $\delta_j=1-\lambda_2(D_j)=1-\omega_j+\omega_j\gamma_+(Q_j)$, this is equivalent to $1-\omega_j\leq\delta_j/16$. Lemma 3 in Appendix~\ref{app:flat} constructs a real polynomial bounded by one on $[-1,1]$, at most $\eta_f$ on $[-1,1-\delta_j]$, and at least $1-\eta_f$ on $[1-\delta_j/16,1]$. The lemma holds for $0<\delta_j<2$, including $\lambda_2(D_j)\leq0$.

In practice, certified bounds $\lambda_2(D_j)\leq a_j<1$ and $\omega_j\geq w_j$, with $a_j>-1$ and $1-w_j\leq(1-a_j)/16$, suffice: use $\delta=1-a_j$. For $0<\eta_f\leq1/2$, the degree is $O((1-a_j)^{-1/2}\log(1/\eta_f))$. For exact spectral endpoints, or enclosures with $1-a_j$ within a fixed factor of $1-\lambda_2(D_j)$, this gives
\begin{equation*}
d_j=O\!\left(
\frac{1}{\sqrt{\gamma_+(Q_j)}}
\log\frac{1}{\eta_f}
\right).
\end{equation*}

For looser spectral bounds, the degree is governed by $1-a_j$.

Let $\square_j=\square_{P^j}$ and
\begin{equation}
W_j=(2\square_j\square_j^\dagger-I)S
\label{eq:quantum-walk}
\end{equation}
be the corresponding Szegedy walk. An eigenvalue $\lambda=\cos\theta$ of $D_j$ with $|\lambda|<1$ becomes the conjugate walk-eigenphase pair $e^{\pm i\theta}$. If $\Lambda_v$ projects onto the complete pair associated with a discriminant eigenvector $|v\rangle$, then
\begin{equation}
\square_j^\dagger\Lambda_v\square_j
=
|v\rangle\langle v|.
\label{eq:compressed-eigenprojector}
\end{equation}

\begin{figure*}[t]
\includegraphics[width=\textwidth]{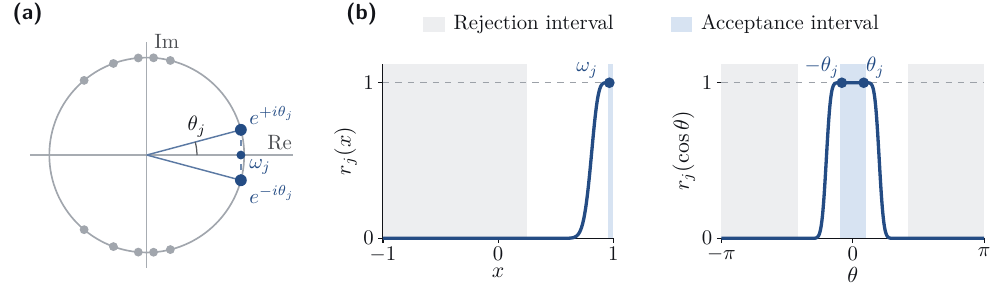}
\caption{Conjugate-eigenphase selection and bounded-Laurent filtering. (a) The blue points $e^{\pm i\theta_j}$ correspond to the Perron eigenvalue $\omega_j=\cos\theta_j$; gray points represent illustrative nontarget eigenphases. (b) The unscaled interval filter $r_j(x)$ from Eq.~(\ref{eq:D-filter-construction}) (left) and its Laurent lift $R_j(e^{i\theta})=r_j(\cos\theta)$ (right). Gray and blue shading indicate the rejection interval $[-1,1-\delta]$ and the acceptance interval $[1-\delta/16,1]$, respectively, and their preimages under $x=\cos\theta$. The illustrative parameters are $\delta=3/4$, $\eta_f=0.05$, $m=2$, $h=25$, and $\theta_j=15^\circ$. Both Perron eigenphases lie in the acceptance interval around zero, with $r_j(1)=1$ and $|R_j|\leq1$.}
\label{fig:laurent}
\end{figure*}

Specializing the conjugate-eigenphase construction \cite{Szegedy2004,Sunderhauf2023} to the Perron pair [Fig.~\ref{fig:laurent}(a)] gives the Perron projector upon compression. At the endpoint $\omega_j=1$, the Perron pair reduces to the $+1$ eigenstate in the relevant invariant walk sector, whose projector likewise compresses to $|\mu_j\rangle\langle\mu_j|$; Appendix~\ref{app:flat} gives the endpoint argument explicitly.

The bounded real filter is transferred to the unit circle [Fig.~\ref{fig:laurent}(b)] by
\begin{equation}
R_j(z)
=
r_j\!\left(\frac{z+z^{-1}}{2}\right).
\label{eq:laurent-filter}
\end{equation}

For $z=e^{i\theta}$,
\begin{equation}
R_j(e^{i\theta})=r_j(\cos\theta),
\qquad
\sup_{|z|=1}|R_j(z)|\leq1.
\label{eq:laurent-filter-bound}
\end{equation}

This Laurent representation provides a GQSP realization of the spectral filter without changing its asymptotic gap dependence.

Multiplying by $z^{d_j}$ gives an ordinary polynomial of degree at most $2d_j$, implemented with one GQSP signal qubit \cite{MotlaghWiebe2024}. Its accepted amplitude operator is $A_j=W_j^{d_j}r_j(X_j)$, where $X_j=(W_j+W_j^\dagger)/2$, and its positive response is $A_j^\dagger A_j=r_j(X_j)^2$. The compensating delay commutes with the signal phase and cancels exactly between filtering and uncomputation. Thus one selective phase uses at most $4d_j$ controlled walk applications, with no unconditional delay, and $8d_j+2$ clean powered-transition calls including the outer preparation and uncomputation. An optional margin $\widetilde r_j=(1-\kappa_j)r_j$ gives $\eta_{\phi,j}^{\rm filt}\leq2\sqrt{2(\kappa_j+\eta_f)}$, so $\kappa_j,\eta_f=O(\eta_\phi^2)$ suffices.

Unlike generalized quantum eigenvalue transform (GQET) amplitude encoding \cite{Claudon2025}, the selective phase requires only the positive response. The Laurent shift cancels from this response, so a bounded real filter needs no further normalization. This can increase the signal-polynomial degree relative to GQET, which also has unit-normalized constructions for suitable filters. Appendix~\ref{app:flat} compares the transformation cores; neither construction is universally cheaper.

\textbf{Proposition 2 (Flat-discriminant selective phase).}
Under the filtering hypotheses and spectral-enclosure condition above, a selective phase about $|\mu_j\rangle$ can be implemented with one GQSP signal qubit and
\begin{equation*}
O\!\left(
\frac{1}{\sqrt{\gamma_+(Q_j)}}
\log\frac{1}{\eta_\phi}
\right)
\end{equation*}
walk applications. If $c_{P^j}$ is the primitive cost of a clean coherent transition oracle for $P^j$, the primitive-oracle cost is
\begin{equation*}
O\!\left(
\frac{c_{P^j}}{\sqrt{\gamma_+(Q_j)}}
\log\frac{1}{\eta_\phi}
\right).
\end{equation*}

For example, consider a four-state cycle with holding probability $\alpha$ and clockwise and counterclockwise probabilities $5(1-\alpha)/8$ and $3(1-\alpha)/8$. Its flat eigenvalues are $\omega=\alpha+(1-\alpha)\sqrt{15}/4$, $\alpha$ (twice), and $\alpha-(1-\alpha)\sqrt{15}/4$. For sufficiently small $\alpha>0$,
\begin{equation*}
\gamma_+(Q)=1-\alpha/\omega,\qquad
\gamma_{\rm abs}(Q)=2\alpha/\omega.
\end{equation*}
Here $1-\omega\leq(1-\alpha)/16$, so the filter premise holds. As $\alpha\to0^+$, the Poincar\'e gap tends to one but the absolute gap tends to zero. The filter degree remains $O(\log(1/\eta_\phi))$: a negative eigenvalue does not obstruct leading-eigenvalue selection. This is not a preparation speedup, since the stationary QSAMPLE is uniform.

To relate the selected Perron state to the stationary sample, define
\begin{equation}
\Delta_j=1-\langle\pi|D_j|\pi\rangle,
\label{eq:flat-defect}
\end{equation}
and $n_j=\omega_j-\langle\pi|D_j|\pi\rangle$. The bias satisfies
\begin{equation}
\begin{aligned}
\frac{n_j}{\omega_j-\lambda_{\min}(D_j)}
&\leq b_j^2\leq\frac{n_j}{\omega_j-\lambda_2(D_j)}\\
&\leq\frac{\Delta_j}{\gamma_+(Q_j)}.
\end{aligned}
\label{eq:flat-bias-bound}
\end{equation}

Since $D_j$ is a contraction, the lower bound gives $b_j\geq\sqrt{n_j/2}$. The sharper upper bound is useful when the small numerator can be enclosed accurately. The coarser bound requires only a deficit and gap certificate. The regime $\Delta_j\ll\gamma_+(Q_j)$ is the Poincar\'e-gap analogue of reversibility on $\pi$-average \cite{Claudon2025}. Appendix~\ref{app:flat} gives the proof and the full filter analysis.

\subsection{Stationary-state selection with forward and reverse access}
\label{subsec:curved-route}

For a Markov kernel $P$ with stationary distribution $\pi$, its time reversal is
\begin{equation*}
P^\star(x,y)
=
\frac{\pi(y)}{\pi(x)}P(y,x).
\end{equation*}

The curved discriminant is
\begin{equation*}
D_{\mathrm{curv}}
=
D(P,P^\star)
=
M_\pi^{1/2}PM_\pi^{-1/2},
\end{equation*}
where $M_\pi=\operatorname{diag}(\pi(x))$. The stationary QSAMPLE $|\pi\rangle$ is a left and right singular vector with singular value one. For a positive integer $k$, define
\begin{equation*}
A_k
:=
D(P^k,(P^\star)^k),
\qquad
\gamma_k(P)
=
\gamma\!\left(P^k(P^\star)^k\right).
\end{equation*}

Here $\gamma$ denotes the absolute spectral gap of the reversible kernel in its argument. The singular-value filter of Ref.~\cite{Claudon2025} has degree
\begin{equation}
d_k
=
O\!\left(
\frac{1}{\sqrt{\gamma_k(P)}}
\log\frac{1}{\eta_f}
\right),
\label{eq:curved-filter-degree}
\end{equation}
accepts the stationary singular component with unit response, and suppresses the others.

We apply Hermitianization \cite{Gilyen2019,Sunderhauf2023} to the forward/reverse transition encodings, with block-index qubit $b$, and denote the resulting accepted filter operator by $F$. For either polynomial parity, we obtain a block-diagonal squared response $F^\dagger F$. This is the relevant object because the selective-phase interface depends on the positive response rather than on the parity-dependent form of $F$ itself. Its stationary target support is
\begin{equation}
\Lambda_\pi^{(b)}
=
I_b\otimes\Pi_\pi.
\label{eq:curved-target-projector}
\end{equation}

Applying the phase to the complete projected-filter success subspace gives the common clean-register selective phase with error at most $2\eta_f$.

\textbf{Proposition 3 (Curved-discriminant selective phase).}
For an ergodic kernel with coherent forward and time-reversed powered-transition access and the curved-filter spectral assumptions of Ref.~\cite{Claudon2025}, the stationary-state selective phase has primitive-oracle cost
\begin{equation}
O\!\left(
\frac{c_{P^k}+c_{(P^\star)^k}}
{\sqrt{\gamma_k(P)}}
\log\frac{1}{\eta_\phi}
\right).
\label{eq:curved-phase-cost}
\end{equation}

With $\gamma_\infty(P)=\max_{k\geq1}\gamma_k(P)/k$ and a maximizing $k_{\mathrm{rev}}$, if both powered-oracle costs are $O(k_{\mathrm{rev}}c_P)$, Eq.~\eqref{eq:curved-phase-cost} becomes $O(c_P\sqrt{k_{\mathrm{rev}}/\gamma_\infty(P)}\log(1/\eta_\phi))$.
Appendix~\ref{app:curved} illustrates the Hermitianized encoding and parity structure and establishes the complete projected-filter phase realization.

At schedule endpoint $i$, the two routes therefore enter Theorem 1 through the cost factors
\begin{equation}
C_i^{\mathrm{flat}}
=
\frac{c_{P_i^{j_i}}}
{\sqrt{\gamma_+(Q_{i,j_i})}},
\qquad
C_i^{\mathrm{curv}}
=
\frac{
c_{P_i^{k_i}}+
c_{(P_i^\star)^{k_i}}
}
{\sqrt{\gamma_{k_i}(P_i)}}.
\label{eq:route-phase-costs}
\end{equation}

Table~\ref{tab:resource-comparison} gives the operational distinction between the two routes. These factors describe one selective phase; the complete preparation also includes the overlap-dependent FPAA lengths and initialization.

\begin{table*}[t]
\caption{Operational comparison of the two nonreversible routes. Phase costs omit accuracy-dependent logarithms and use spectral bounds of the precision specified above. The spectral constructions are from Ref.~\cite{Claudon2025}; the common-embedding FPAA composition follows Ref.~\cite{Zhao2026}.}
\label{tab:resource-comparison}
\begin{ruledtabular}
\begin{tabular}{p{0.19\textwidth}p{0.34\textwidth}p{0.39\textwidth}}
Quantity & Flat route & Curved route \\
\hline
Required clean access & $P_i^{j_i}$, inverse, and controlled versions & $P_i^{k_i}$ and $(P_i^\star)^{k_i}$, inverses, and controlled versions \\
Selected state & Perron state $|\mu_i\rangle$ & Stationary state $|\pi_i\rangle$ \\
Spectral input & Primitive discriminant, gap $\gamma_+(Q_{i,j_i})$, and acceptance-interval condition & Powered singular gap $\gamma_{k_i}(P_i)>0$ \\
Leading phase cost & $c_{P_i^{j_i}}/\sqrt{\gamma_+(Q_{i,j_i})}$ & $(c_{P_i^{k_i}}+c_{(P_i^\star)^{k_i}})/\sqrt{\gamma_{k_i}(P_i)}$ \\
Stationary-state bias & $b_i$, bounded by Eq.~\eqref{eq:flat-bias-bound} & Zero \\
Transport & FPAA along Perron states with certified $\underline q_i$; endpoint-bias allowance & FPAA along stationary states with certified $\underline p_i$ \\
\end{tabular}
\end{ruledtabular}
\end{table*}

The explicit curved Hermitianization uses one additional block-index qubit; direct generalized quantum singular value transformation (GQSVT) instead acts on the right-singular space without this dilation qubit, with implementation-dependent resources. Route-specific transport bounds are given in Appendix~\ref{app:transport}.

\section{Resources and parameter certification}
\label{sec:resources}

\subsection{Implementation cost}
\label{subsec:implementation-cost}

Let $N=|\Omega|$, $n=\lceil\log_2N\rceil$, and let $s_i$ denote the arithmetic precision at endpoint $i$. Let $G_i(s_i)$ upper-bound the elementary reversible-gate cost of the required controlled clean powered-transition oracles and inverses, including scratch-register uncomputation. Define
\begin{equation}
g_i=
\begin{cases}
\gamma_+(Q_{i,j_i}), & \text{flat route}\\
\gamma_{k_i}(P_i), & \text{curved route}
\end{cases}
\label{eq:route-gap}
\end{equation}

Let $\eta_{\mathrm{ph}}$ denote the target error of a compiled selective phase, and let $t_i$ denote the synthesis precision allocated to each required filter rotation at endpoint $i$. The corresponding filter degree is
\begin{equation}
d_i=
O\!\left(
\frac{1}{\sqrt{g_i}}
\log\frac{1}{\eta_{\mathrm{ph}}}
\right).
\label{eq:filter-degree}
\end{equation}

The stated gap scaling assumes certified spectral bounds within a fixed factor of the corresponding gaps. With looser enclosures, the cost is determined by the supplied bounds, using $1-a_j$ for the flat filter.

Let $R_{\mathrm{rot}}(t_i)$ be the cost of synthesizing such a rotation to precision $t_i$. A selective phase at endpoint $i$ can be bounded by
\begin{equation}
\mathcal G_i
=O\!\left(
(d_i+1)
\left[
G_i(s_i)+O(n)+R_{\mathrm{rot}}(t_i)
\right]\right).
\label{eq:endpoint-gate-cost}
\end{equation}

The terms respectively account for the controlled powered-oracle implementation, structural walk and reflection operations, and filter-rotation synthesis; the precision parameters are chosen so that their accumulated errors fit the phase budget.

If $G_{\mathrm{init}}$ is the gate cost of the initial state, the full transport gate complexity is
\begin{equation}
G_{\mathrm{total}}
=
O\!\left(
G_{\mathrm{init}}
+
\sum_{i=0}^{\ell-1}
L_i\bigl[\mathcal G_i+\mathcal G_{i+1}\bigr]
\right).
\label{eq:total-gate-cost}
\end{equation}

The corresponding total-space count is
$Q_{\mathrm{total}}=2n+Q_{\mathrm{struct}}+Q_{\mathrm{oracle}}(s)+Q_{\mathrm{arith}}(s)$,
for the state registers, structural qubits, oracle workspace, and arithmetic workspace, respectively (Appendix~\ref{app:resources}). Structured Gibbs transitions can supply the required oracle inputs through local energy differences. Let $E$ be the energy function and $\beta$ the inverse temperature, so that $\pi_\beta(x)\propto e^{-\beta E(x)}$. The ratio $\pi_\beta(y)/\pi_\beta(x)=e^{-\beta[E(y)-E(x)]}$ removes the partition function from a Metropolis--Hastings acceptance step.  The controlled-oracle realization, precision allocation, and complete gate and workspace accounting are given in Appendix~\ref{app:resources}.

\subsection{Parameter certificates}
\label{subsec:parameter-certificates}

For a Gibbs schedule with energy range $E_{\max}-E_{\min}\leq B$, adjacent stationary QSAMPLEs satisfy
\begin{equation}
p_i
\geq
\operatorname{sech}^{2}\!\left(\frac{B|\beta_{i+1}-\beta_i|}{4}\right).
\label{eq:gibbs-overlap-bound}
\end{equation}

This bound has quadratic small-step behavior and gives certified stationary-state overlaps without evaluating the partition function. Appendix~\ref{app:certificates} gives its proof and the corresponding step-size condition.

For the flat route, let $D_i=D(P_i^{j_i},P_i^{j_i})$, define
\begin{equation*}
\gamma_i
:=
\gamma_+(Q_{i,j_i}),
\qquad
\Delta_i
:=
1-\langle\pi_i|D_i|\pi_i\rangle,
\end{equation*}
and fix constants $0<q_{\min}<p_{\min}\leq1$ with
\begin{equation}
b_\star
:=
\sin\!\left(
\frac{\arccos\sqrt{q_{\min}}-\arccos\sqrt{p_{\min}}}{2}
\right).
\label{eq:bias-threshold}
\end{equation}

\textbf{Lemma 1 (Sufficient condition for quantitative flat-route overlaps).}
Suppose each powered flat discriminant is primitive and satisfies the filtering hypotheses, the stationary schedule obeys $p_i\geq p_{\min}$, and certified bias bounds satisfy $b_i\leq\overline b_i\leq b_\star$. Then
\begin{equation}
q_i\geq q_{\min}>0
\label{eq:certified-min-overlap}
\end{equation}
throughout the sequence. A sufficient, coarser condition is
\begin{equation}
\frac{\Delta_i}{\gamma_i}\leq b_\star^2.
\label{eq:flat-defect-threshold}
\end{equation}

Primitivity already makes all Perron vectors strictly positive and their overlaps nonzero. The certificate supplies the quantitative lower bounds needed to control amplification cost. Stage-dependent bounds, taking the smaller justified bias estimate in Eq.~\eqref{eq:flat-bias-bound}, can instead be inserted directly in Eq.~\eqref{eq:overlap-bound}. The initial and final biases must separately fit Eq.~\eqref{eq:endpoint-error}.

\textbf{Lemma 2 (Direct Perron-state overlap certificate).}
Let $D_i$ and $D_{i+1}$ be primitive flat discriminants
with certified spectral bounds
$\omega_i\geq w_i$ and $\lambda_2(D_i)\leq a_i$
at each stage. Define
\begin{equation}
g_i^{\mathrm{cross}}
=
\max\{w_i-a_{i+1},w_{i+1}-a_i\}.
\end{equation}

If $\overline e_i\geq\|D_{i+1}-D_i\|$ and
$g_i^{\mathrm{cross}}>0$, then
\begin{equation}
q_i\geq
\max\left\{
0,1-\frac{\overline e_i^2}
{(g_i^{\mathrm{cross}})^2}
\right\}.
\label{eq:direct-overlap-bound}
\end{equation}

Here the norm is the spectral operator norm.
Unlike Eq.~\eqref{eq:overlap-bound}, this certificate
does not require stationary-state overlaps or
stationary-to-Perron bias bounds. Whenever it gives
a positive lower bound, it can be used directly in
Theorem 1. The filtering conditions and the initial
and final bias requirements must still be satisfied.
The proof is given in Appendix~\ref{app:certificates}. This certificate is a rank-one application of the Davis--Kahan $\sin\theta$ perturbation bound, using the larger of the two directional cross-gaps \cite{DavisKahan1970}.

For the curved route, if the second-largest singular value of $D(P_i^{k_i},(P_i^\star)^{k_i})$ is at most $r_i<1$, then $\gamma_{k_i}(P_i)\geq1-r_i^2$. The Gibbs overlap derivation, proof of Lemma 1, and route-specific certification conditions are given in Appendix~\ref{app:certificates}.

\section{Finite-dimensional example}
\label{sec:verification}

We evaluate the transport quantities in an eight-state nonreversible Gibbs family on $\mathbb Z_4\times\{+,-\}$. The dynamics combine a velocity-dependent Metropolis--Hastings kernel $Q_\beta$ with a reversible velocity transition $R$, giving $P_\beta=Q_\beta R$ and $P_\beta^\star=RQ_\beta$. The energy range is $B=1.2$, and the schedule is $\beta=0,0.75,1.25$. The transition probabilities and spectral certificates are given in Appendix~\ref{app:verification}.

Since intermediate biases affect both the transport overlaps and the spectral filtering costs, Table~\ref{tab:certified-comparison} compares preparation paths at fixed final stationary-state accuracy. All paths start from the exactly prepared uniform state, whose selective phase is implemented directly. The staged flat paths use $j_0=j_1=4$ or $j_0=j_1=1$ and the same final power-$4$ Perron target. Both initial powers select the uniform state. The direct flat path omits the intermediate stage, while the direct curved path uses power $k=1$ and requires both forward and reverse access.

\begin{table*}[t]
\caption{Preparation at common final trace-distance target $10^{-3}$. Degrees $d$ refer to spectrally selected endpoints; the initial phase uses no filter. $L$ is the odd FPAA parameter. $N_O$ counts clean powered row-oracle calls, including controls, inverses, and outer preparation/uncomputation. The first two staged rows use stationary geometry; the next two use direct overlap certificates. These are construction counts under the table model of Appendix~\ref{app:resources}, not optimized gate costs. Overlap bounds are rounded down and error bounds up.}
\label{tab:certified-comparison}
\begin{ruledtabular}
\begin{tabular}{lccccc}
Construction & $\underline q_i$ & $d$ & $L$ & $N_O$ & Final error $\leq$\\
\hline
Staged flat, $j_1=4$ & $(0.9510,0.9778)$ & $(52,52)$ & $(5,5)$ & $2508$ & $9.66\times10^{-4}$\\
Staged flat, $j_1=1$ & $(0.9478,0.9756)$ & $(108,52)$ & $(5,5)$ & $4300$ & $9.66\times10^{-4}$\\
Staged flat, $j_1=4$, Lemma 2 & $(0.969514,0.986234)$ & $(52,52)$ & $(5,5)$ & $2508$ & $9.66\times10^{-4}$\\
Staged flat, $j_1=1$, Lemma 2 & $(0.875781,0.766915)$ & $(116,52)$ & $(7,7)$ & $6834$ & $9.66\times10^{-4}$\\
Direct flat, $j=4$ & $0.8715$ & $44$ & $5$ & $708$ & $9.66\times10^{-4}$\\
Direct curved, $k=1$ & $0.8715$ & $10$ & $5$ & $160$ & $9.0\times10^{-4}$\\
\end{tabular}
\end{ruledtabular}
\end{table*}

We reserve $10^{-4}$ for the flat endpoint mismatch, $4.5\times10^{-4}$ for total intrinsic FPAA error, and $4.5\times10^{-4}$ for accumulated phase error. The certified final flat bias is at most $6.54\times10^{-5}$. The same algorithmic allowances suffice for the zero-bias curved target. Appendix~\ref{app:verification} supplies interval certificates, explicit filters, and implementation tolerances.

\begin{table*}[t]
\caption{Adjacent squared-overlap certificates for the same two staged paths. Each pair gives the first and second step. The first three rows are lower bounds; the last row gives numerical estimates.}
\label{tab:overlap-comparison}
\begin{ruledtabular}
\begin{tabular}{lcc}
Certificate & $j_1=1$ & $j_1=4$\\
\hline
Geometry, energy range & $(0.9478,0.9756)$ & $(0.9510,0.9778)$\\
Geometry, tight stationary overlap & $(0.970469,0.987547)$ & $(0.972851,0.989089)$\\
Direct (Lemma 2) & $(0.875781,0.766915)$ & $(0.969514,0.986234)$\\
Numerical overlap & $(0.971034,0.990243)$ & $(0.972856,0.989104)$\\
\end{tabular}
\end{ruledtabular}
\end{table*}

Table~\ref{tab:overlap-comparison} compares the overlap certificates. Tight stationary-state geometry gives the strongest bounds, while Lemma 2 improves upon the energy-range certificate for $j_1=4$ using only discriminant differences and spectral bounds. As Table~\ref{tab:certified-comparison} shows, the direct certificate increases the construction cost for $j_1=1$ but not for $j_1=4$, without changing the final error bound. The benefit is reduced information requirements, not a speedup.

\begin{figure}[t]
\includegraphics[width=\columnwidth]{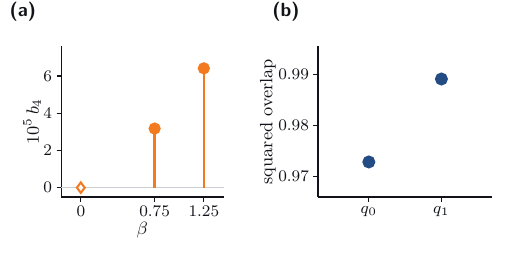}
\caption{Flat-target bias and transport overlaps in the eight-state Gibbs family ($j=4$). (a) Stationary-to-Perron trace-distance bias, plotted as $10^5b_4(\beta)$. The scaled biases are $10^5b_4(0.75)=3.18$ and $10^5b_4(1.25)=6.42$. The open diamond marks the exact value $b_4(0)=0$. (b) Selected-state squared overlaps $q_0=0.972856$ for $\beta:0\to0.75$ and $q_1=0.989104$ for $0.75\to1.25$. These are numerical estimates, not certified lower bounds. The corresponding stationary-state overlaps are $p_0=0.972862$ and $p_1=0.989110$.}
\label{fig:gibbs-schedule}
\end{figure}

Figure~\ref{fig:gibbs-schedule} shows the numerical biases and overlaps for the power-$4$ schedule. Reducing the intermediate power to $j=1$ raises the estimated bias from $3.18\times10^{-5}$ to $6.66\times10^{-3}$. The certified lower bound $b_1(0.75)\geq3.614\times10^{-3}$ exceeds the final tolerance $10^{-3}$. This bias is a diagnostic, not an input to the direct certificate. Both selectors satisfy the filter premise, but the smaller power has weaker spectral separation and requires a higher degree. The staged cost therefore increases. Direct flat transport is cheaper because its initial-to-final overlap remains large; the curved baseline has the smallest bound but also requires reverse access. In the table model, changing the power does not change the cost of a row-oracle call.

Writing $c_j$ for the cost of a power-$j$ oracle, the two geometry-certified staged constructions have primitive costs $3464c_1+836c_4$ and $2508c_4$. The more biased intermediate target is therefore preferable between these two staged paths when $c_4/c_1>866/418\simeq2.07$. Direct flat preparation remains cheaper in this example.

A separate winning-streak example in Appendix~\ref{app:verification} illustrates the distinction between worst-start mixing, spectral separation, target bias, and initialization overlap. Its nonprimitive regime lies outside the hypotheses of the flat-route theorem; the query proxy describes only part of the preparation cost.

\subsection{Uniform certification over a temperature interval}

The same model gives a certificate for a continuous range of targets. Write $D_\beta=D(P_\beta,P_\beta)$ for $0\leq\beta\leq0.75$. The transition rules imply
\begin{equation*}
\|D_\beta-D_{\beta'}\|\leq\tfrac54|\beta-\beta'|.
\end{equation*}
At the 26 anchor points $\beta_k=0.03k$, $k=0,\ldots,25$, we certify $\omega(\beta_k)\geq0.991$ and $\lambda_2(D_{\beta_k})\leq0.501$. Every temperature lies within $0.015$ of an anchor. Weyl's inequality gives uniform bounds $\omega(\beta)\geq0.97225$ and $\lambda_2(D_\beta)\leq0.51975$, which satisfy the filter premise. Lemma 2 then gives
\begin{equation}
q(\beta,\beta')\geq1-\left(\frac{1.25|\beta-\beta'|}{0.4525}\right)^2.
\label{eq:uniform-overlap}
\end{equation}
Thus any schedule in the interval with steps at most $0.03$ has adjacent squared overlaps at least $0.9931$. Intermediate stationary distributions and biases are not needed, and the schedule can be refined without certifying each new spectrum. Appendix~\ref{app:verification} gives the proof and anchor checks. This finite example does not establish an asymptotic reduction in certification cost.

\section{Discussion and conclusion}
\label{sec:discussion}

The coherent-state preparation problem considered here differs from quantum-enhanced MCMC with quantum proposals and classical acceptance steps \cite{Layden2023}, for which limitations on mixing speedup have been established \cite{OrfiSels2024}.

It also differs from thermal-state preparation for general quantum Hamiltonians using quantum detailed-balance dynamics \cite{ChenKastoryanoGilyenExact2025,ChenKastoryanoNature2025,Ding2025}.

The framework separates the spectral requirements for intermediate transport from the stationary-state accuracy requirements at the endpoints. Direct discriminant certificates determine transport parameters without intermediate stationary-state or bias estimates. The Gibbs example demonstrates this even when an intermediate bias exceeds the final tolerance, and uniform bounds allow a temperature schedule to be refined within a certified interval.

Gate-level costs depend on clean powered-transition oracles and finite-precision operations. Direct preparation remains cheaper in the finite example; we establish no general resource advantage over reversible alternatives or other preparation methods. Identifying structured families in which this flexibility also reduces preparation cost remains open.

\section*{Data Availability}

The numerical results presented in this work can be reproduced from the equations, transition rules, parameters, and computational procedures specified in the manuscript, particularly in Appendix~\ref{app:verification}. No experimental or external datasets were used.

\section*{AI Use Disclosure}

OpenAI ChatGPT (GPT-6 and GPT-6.2) and Anthropic Claude (Opus 5.5) assisted with mathematical derivations, numerical calculations and verification, literature review, manuscript writing, and LaTeX formatting. The author directed the calculations, reviewed the derivations and numerical certificates, and checked the AI-assisted outputs against the stated mathematical and computational procedures. The author is responsible for the manuscript and its scientific conclusions.

\appendix
\input{appendix_A}
\input{appendix_B}
\input{appendix_C}
\input{appendix_D}
\input{appendix_E}
\input{appendix_F}
\input{appendix_G}
\input{appendix_H}

\bibliography{references}

\end{document}

%% file: appendix_A.tex
\section{Coherent access and discriminant encodings}
\label{app:access}

\subsection{QSAMPLE states and accuracy conventions}

Let $\Omega$ be a finite state space of size $N$, and let $P$ be an ergodic row-stochastic transition matrix,
\begin{equation*}
P(x,y)\geq0,
\qquad
\sum_{y\in\Omega}P(x,y)=1.
\end{equation*}

Its unique stationary distribution $\pi$ satisfies
\begin{equation*}
\sum_x\pi(x)P(x,y)=\pi(y).
\end{equation*}

We assume full stationary support, $\pi(x)>0$, as follows from finite-state ergodicity. The coherent amplitude encoding of $\pi$, called its quantum sample or QSAMPLE, is
\begin{equation*}
|\pi\rangle
=
\sum_{x\in\Omega}\sqrt{\pi(x)}\,|x\rangle,
\qquad
\langle\pi|\pi\rangle=1.
\end{equation*}

For any probability distribution $p$, the notation $|p\rangle$ denotes the analogous normalized amplitude vector. The rank-one projector onto the stationary QSAMPLE is
\begin{equation*}
\Pi_\pi=|\pi\rangle\langle\pi|.
\end{equation*}

We measure quantum-state accuracy using trace distance,
\begin{equation*}
D_{\mathrm{tr}}(\rho,\sigma)
=
\frac12\|\rho-\sigma\|_1.
\end{equation*}

A state $\rho$ is an $\epsilon$-accurate QSAMPLE of $\pi$ if
\begin{equation*}
D_{\mathrm{tr}}(\rho,\Pi_\pi)\leq\epsilon.
\end{equation*}

For normalized pure states,
\begin{equation}
\label{eq:A7}
D_{\mathrm{tr}}
\bigl(
|\psi\rangle\langle\psi|,
|\varphi\rangle\langle\varphi|
\bigr)
=
\sqrt{1-|\langle\psi|\varphi\rangle|^2}.
\end{equation}

The trace-distance criterion also controls the measurement distribution. Let $\mathcal M$ be computational-basis measurement. Since trace distance is contractive under completely positive trace-preserving maps,
\begin{equation*}
D_{\mathrm{tr}}\bigl(\mathcal M(\rho),\mathcal M(\Pi_\pi)\bigr)
\leq D_{\mathrm{tr}}(\rho,\Pi_\pi).
\end{equation*}

The measured distribution of $\Pi_\pi$ is $\pi$. Identifying trace distance between classical diagonal states with total-variation distance gives
\begin{equation}
\label{eq:A9}
d_{\mathrm{TV}}(p_\rho,\pi)
\leq D_{\mathrm{tr}}(\rho,\Pi_\pi),
\end{equation}
where $p_\rho$ is the computational-basis distribution obtained from $\rho$.

\subsection{Transition-state isometries and discriminant compression}

Set $n=\lceil\log_2N\rceil$. A coherent transition-state oracle for $P$ is a unitary $O_P$ satisfying
\begin{equation*}
O_P|x\rangle|0^n\rangle
=
|x\rangle
\sum_{y\in\Omega}
\sqrt{P(x,y)}\,|y\rangle.
\end{equation*}

For $N\neq2^n$, the oracle extension preserves the designated subspace of valid state labels. The action on other register states is completed by a unitary extension. We assume access to this unitary, its inverse, and the controlled versions used by the chosen spectral circuit. Any arithmetic scratch registers return to zero. A classical sampler or a preparation retaining a proposal or path label does not alone supply this clean oracle.

Restricting the oracle to inputs with a clean second register defines
\begin{equation}
\label{eq:A11}
\square_P:
|x\rangle
\longmapsto
|x\rangle
\sum_y\sqrt{P(x,y)}\,|y\rangle.
\end{equation}

For basis states $x,x'$,
\begin{equation*}
\begin{aligned}
\langle x'|\square_P^\dagger\square_P|x\rangle
&=
\delta_{x,x'}
\sum_yP(x,y)\\
&=\delta_{x,x'}.
\end{aligned}
\end{equation*}

Thus $\square_P$ is an isometry: $\square_P^\dagger\square_P=I$.

Let
\begin{equation*}
S|x\rangle|y\rangle
=
|y\rangle|x\rangle
\end{equation*}
be the register-swap operator. It satisfies $S=S^\dagger=S^{-1}$.

For any two row-stochastic kernels $P_1,P_2$, define
\begin{equation}
\label{eq:A15}
D(P_1,P_2)
=
\square_{P_1}^\dagger S\square_{P_2}.
\end{equation}

Its matrix elements follow directly:
\begin{equation*}
\begin{aligned}
\langle x|D(P_1,P_2)|y\rangle
&=
\langle \square_{P_1}x|S|\square_{P_2}y\rangle\\
&=
\sqrt{P_1(x,y)P_2(y,x)}.
\end{aligned}
\end{equation*}

The triple $(S,\square_{P_1},\square_{P_2})$ is therefore a projected unitary encoding of the discriminant $D(P_1,P_2)$.

Because $S$ is unitary and the two transition maps are isometries,
\begin{equation*}
\|D(P_1,P_2)\|\leq1.
\end{equation*}

Two choices are central to the paper. Using the same kernel on both sides gives the symmetric flat discriminant,
\begin{equation}
\label{eq:A18}
\begin{aligned}
D_{\mathrm{flat}}(P)
&=
D(P,P),\\
D_{\mathrm{flat}}(P)(x,y)
&=
\sqrt{P(x,y)P(y,x)}.
\end{aligned}
\end{equation}

\subsection{Time-reversed and powered-transition access}

The time reversal of $P$ with respect to $\pi$ is
\begin{equation}
\label{eq:A19}
P^\star(x,y)
=
\frac{\pi(y)}{\pi(x)}P(y,x).
\end{equation}

Stationarity implies
\begin{equation*}
\sum_yP^\star(x,y)
=
\frac{1}{\pi(x)}
\sum_y\pi(y)P(y,x)
=1.
\end{equation*}

Thus $P^\star$ is row stochastic. Its stationary distribution is also $\pi$, since
\begin{equation*}
\sum_x\pi(x)P^\star(x,y)
=
\pi(y)\sum_xP(y,x)
=\pi(y).
\end{equation*}

It is the adjoint of $P$ in the complex $\pi$-weighted inner product,
\begin{equation*}
\langle f,g\rangle_\pi
=
\sum_x\pi(x)\overline{f(x)}g(x).
\end{equation*}

Indeed,
\begin{equation*}
\langle f,Pg\rangle_\pi
=
\langle P^\star f,g\rangle_\pi.
\end{equation*}

For a positive integer $m$, a clean coherent oracle for $P^m$ has the action
\begin{equation}
\label{eq:A24}
O_{P^m}|x\rangle|0^n\rangle
=
|x\rangle
\sum_y\sqrt{P^m(x,y)}\,|y\rangle.
\end{equation}

We denote its primitive implementation cost by $c_{P^m}$. The corresponding cost for the time-reversed powered kernel is $c_{(P^\star)^m}$.

A sequence of $m$ coherent one-step transitions generally retains intermediate path information. Producing the clean endpoint amplitudes in Eq.~\eqref{eq:A24} therefore requires an implementation of the powered kernel, including the necessary compression or uncomputation. Accordingly, the relation
\begin{equation}
\label{eq:A25}
c_{P^m}=O(mc_P)
\end{equation}
is used only when a clean powered-oracle construction with that cost is available.

The flat route requires $O_{P^j}$ and its inverse, including controlled calls. The curved route requires these operations for both $P^k$ and $(P^\star)^k$, with coherent selection by a block-index qubit. The inverse $O_P^\dagger$ unprepares a row state; it is not an oracle for $P^\star$. These costs include all controls and scratch uncomputation. Appendix~\ref{app:resources} supplies a finite-table realization and a sufficient structured row-preparation condition.

%% file: appendix_B.tex
\section{Multistage transport and stationary-state accuracy}
\label{app:transport}
We prove Theorem 1 and its stationary-state consequences from Sec.~\ref{sec:transport}.

\subsection{Common clean-register embeddings}

Let $\mathcal H_{\mathrm{sys}}$ be the system state space. After padding the work registers used by the different spectral constructions, let $\mathcal H_{\mathrm{work}}$ be a common implementation space.

Choose an isometry
\begin{equation}
\label{eq:B1}
J:\mathcal H_{\mathrm{sys}}
\longrightarrow\mathcal H_{\mathrm{work}},
\qquad
J^\dagger J=I,
\end{equation}
representing the designated clean initialization of all work registers.

At stage $i$, the spectral construction selects a normalized state $|\sigma_i\rangle$, with projector
\begin{equation*}
\Pi_i=|\sigma_i\rangle\langle\sigma_i|.
\end{equation*}

The corresponding ideal phase on the system is
\begin{equation*}
S_\phi(\Pi_i)
=
I+(e^{i\phi}-1)\Pi_i.
\end{equation*}

Suppose the route-specific spectral realization naturally uses an isometry
\begin{equation}
\label{eq:B4}
V_i:\mathcal H_{\mathrm{sys}}
\longrightarrow\mathcal H_{\mathrm{work}}
\end{equation}
and implements a unitary $\widehat S_{\phi,i}$ satisfying
\begin{equation}
\label{eq:B5}
\left\|
\widehat S_{\phi,i}V_i
-
V_iS_\phi(\Pi_i)
\right\|
\leq\eta_{\phi,i}.
\end{equation}

Since $J$ and $V_i$ are isometries with the same domain and a common codomain, the map $J|\psi\rangle\mapsto V_i|\psi\rangle$ is an isometry between subspaces of equal dimension. Extending orthonormal bases of those subspaces gives a unitary $T_i$ on $\mathcal H_{\mathrm{work}}$ such that
\begin{equation}
\label{eq:B6}
T_iJ=V_i.
\end{equation}

Define the compiled phase in the common embedding by
\begin{equation}
\label{eq:B7}
\widetilde S_{\phi,i}
=
T_i^\dagger\widehat S_{\phi,i}T_i.
\end{equation}

Using Eqs.~\eqref{eq:B5}--\eqref{eq:B7},
\begin{equation}
\label{eq:B8}
\begin{aligned}
\widetilde S_{\phi,i}J-JS_\phi(\Pi_i)
&=
T_i^\dagger
\left[
\widehat S_{\phi,i}V_i
-
V_iS_\phi(\Pi_i)
\right].
\end{aligned}
\end{equation}

Unitarity therefore gives
\begin{equation}
\label{eq:B9}
\left\|
\widetilde S_{\phi,i}J
-
JS_\phi(\Pi_i)
\right\|
\leq\eta_{\phi,i}.
\end{equation}

The preparation and uncomputation represented by $T_i$ and $T_i^\dagger$ are included in the cost of $\widetilde S_{\phi,i}$. For the two route-specific corollaries these maps have explicit efficient realizations. In the flat route, after fixed register padding choose
\begin{equation*}
J_{\mathrm{flat}}|\psi\rangle
=
|0\rangle_s|\psi\rangle_A|0^n\rangle_B,
\end{equation*}
and
\begin{equation*}
V_i^{\mathrm{flat}}|\psi\rangle
=
|0\rangle_s\square_{P_i^{j_i}}|\psi\rangle.
\end{equation*}

Then
\begin{equation*}
T_i^{\mathrm{flat}}
=
I_s\otimes O_{P_i^{j_i}}
\end{equation*}
satisfies $T_i^{\mathrm{flat}}J_{\mathrm{flat}}=V_i^{\mathrm{flat}}$. Its preparation and unpreparation require only a constant number of calls to the clean powered-transition oracle and are absorbed into $C_i^{\mathrm{flat}}$.

For the curved route, use the common padded clean embedding
\begin{equation*}
J_{\mathrm{curv}}|\psi\rangle
=
|0\rangle_s|0\rangle_b|\psi\rangle_A|0^n\rangle_B.
\end{equation*}

The projected-filter realization of Appendix~\ref{app:curved} already acts in this embedding, so one may take $V_i^{\mathrm{curv}}=J_{\mathrm{curv}}$ and $T_i^{\mathrm{curv}}=I$; the stage-dependent forward and reverse powered-transition preparations occur inside the compiled phase circuit and are included in $C_i^{\mathrm{curv}}$. 

The ideal lifted operation preserves $\operatorname{Ran}(J)$, because
\begin{equation*}
JS_\phi(\Pi_i)|\psi\rangle
\in\operatorname{Ran}(J)
\end{equation*}
for every system state $|\psi\rangle$.

\subsection{Proof of the multistage transport theorem}

Consider normalized selected states
\begin{equation*}
|\sigma_0\rangle,\ldots,|\sigma_\ell\rangle,
\end{equation*}
with adjacent squared overlaps
\begin{equation}
\label{eq:B12}
q_i
=
|\langle\sigma_i|\sigma_{i+1}\rangle|^2>0.
\end{equation}

Let certified positive bounds satisfy
\begin{equation}
\label{eq:B13}
0<\underline q_i\leq q_i.
\end{equation}

Fixed-point amplitude amplification builds on quantum search \cite{Grover1996} and its fixed-point variant \cite{Grover2005}.

The optimal FPAA construction of Yoder, Low, and Chuang \cite{YoderLowChuang2014}, in the multistage setting used by Zhao \cite{Zhao2026}, supplies a sequence parameterized by an odd integer
\begin{equation}
\label{eq:B14}
L_i
=
O\!\left(
\frac{1}{\sqrt{\underline q_i}}
\log\frac{1}{\epsilon_i^{\mathrm{FP}}}
\right),
\end{equation}
which maps the ideal input $|\sigma_i\rangle$ to within trace distance $\epsilon_i^{\mathrm{FP}}$ of $|\sigma_{i+1}\rangle$. More explicitly, for $0<\underline q_i<1$, it suffices to take an odd integer
\begin{equation}
\label{eq:explicit-fpaa-length}
L_i\geq
\frac{\operatorname{arcosh}(1/\epsilon_i^{\mathrm{FP}})}
{\operatorname{artanh}\sqrt{\underline q_i}}.
\end{equation}

The sequence has $(L_i-1)/2$ phases about each endpoint projector. This follows from the FPAA acceptance threshold $\tanh^2[\operatorname{arcosh}(1/\epsilon_i^{\mathrm{FP}})/L_i]$. If $\underline q_i=1$, the two projectors coincide and no transport is needed.

Let $F_i$ denote the ideal FPAA unitary on the system and $\widetilde F_i$ the corresponding compiled unitary on the common work space. Write the ideal sequence as a product of elementary selective phases and denote the compiled counterparts by replacing each elementary phase with its implementation.

At stage $i$, let $\eta_i^{\mathrm{ph}}$ be a common bound on the errors of the selective phases about both $\Pi_i$ and $\Pi_{i+1}$, uniformly over the phase angles appearing in the FPAA sequence.

For products $U_m\cdots U_1$ and $\widetilde U_m\cdots\widetilde U_1$, the telescoping identity is
\begin{equation}
\label{eq:B15}
\begin{aligned}
&\widetilde U_m\cdots\widetilde U_1
-
U_m\cdots U_1\\
&\quad=
\sum_{r=1}^{m}
\widetilde U_m\cdots\widetilde U_{r+1}
(\widetilde U_r-U_r)\\
&\qquad\qquad\times
U_{r-1}\cdots U_1.
\end{aligned}
\end{equation}

For the transport circuit, every ideal intermediate state beginning in $\operatorname{Ran}(J)$ remains in that subspace. Equation~\eqref{eq:B9} therefore bounds each term of the telescoping sum on the ideal trajectory.

Since the sequence contains $L_i-1$ selective phases, telescoping gives an error of at most $(L_i-1)\eta_i^{\mathrm{ph}}$, and hence the sufficient bound
\begin{equation}
\label{eq:B16}
\|\widetilde F_iJ-JF_i\|
\leq
2(L_i-1)\eta_i^{\mathrm{ph}}.
\end{equation}

Let $\widetilde\rho_i$ be the actual work-space state entering stage $i$, and define
\begin{equation}
\label{eq:B17}
d_i
=
D_{\mathrm{tr}}
\bigl(
\widetilde\rho_i,
J\Pi_iJ^\dagger
\bigr).
\end{equation}

For the initial state $\rho_0$ on the system, the corresponding work-space input is
\begin{equation*}
\widetilde\rho_0
=
J\rho_0J^\dagger.
\end{equation*}

Because an isometric embedding preserves trace distance,
\begin{equation*}
d_0
=
D_{\mathrm{tr}}(\rho_0,\Pi_0)
=:\delta_0.
\end{equation*}

We compare the actual stage output with the ideal target by inserting the compiled evolution of the exact selected input:
\begin{equation*}
\begin{aligned}
d_{i+1}
&=
D_{\mathrm{tr}}
\left(
\widetilde F_i\widetilde\rho_i\widetilde F_i^\dagger,
J\Pi_{i+1}J^\dagger
\right)\\
&\leq
D_{\mathrm{tr}}
\left(
\widetilde F_i\widetilde\rho_i\widetilde F_i^\dagger,
\widetilde F_iJ\Pi_iJ^\dagger\widetilde F_i^\dagger
\right)\\
&\quad+
D_{\mathrm{tr}}
\left(
\widetilde F_iJ\Pi_iJ^\dagger\widetilde F_i^\dagger,
JF_i\Pi_iF_i^\dagger J^\dagger
\right)\\
&\quad+
D_{\mathrm{tr}}
\left(
JF_i\Pi_iF_i^\dagger J^\dagger,
J\Pi_{i+1}J^\dagger
\right).
\end{aligned}
\end{equation*}

The first term equals $d_i$ by unitary invariance of trace distance. The second is bounded by Eq.~\eqref{eq:B16}, using the pure-state trace-distance bound in terms of Euclidean distance. The last term is at most $\epsilon_i^{\mathrm{FP}}$ by the ideal FPAA guarantee.

Consequently,
\begin{equation}
\label{eq:B20}
d_{i+1}
\leq
d_i+
2(L_i-1)\eta_i^{\mathrm{ph}}
+
\epsilon_i^{\mathrm{FP}}.
\end{equation}

Iterating from $i=0$ to $i=\ell-1$ gives
\begin{equation}
\label{eq:B21}
D_{\mathrm{tr}}(\rho_\ell,\Pi_\ell)
\leq
\delta_0+
\sum_{i=0}^{\ell-1}
\left[
2(L_i-1)\eta_i^{\mathrm{ph}}
+
\epsilon_i^{\mathrm{FP}}
\right].
\end{equation}

Here $\rho_\ell$ denotes the recovered system output. More explicitly, complete the isometry $J$ to a unitary clean-register embedding and let $\mathcal R$ be the corresponding recovery channel that applies its inverse and discards the work registers. It satisfies
\begin{equation*}
\mathcal R(J\rho J^\dagger)=\rho
\end{equation*}
for every system state $\rho$. The bound is first established against $J\Pi_\ell J^\dagger$ in the common work space; contractivity of trace distance under the completely positive trace-preserving (CPTP) map $\mathcal R$ then gives the stated bound for the recovered system output.

Let $C_i$ be the primitive-oracle cost of one implemented selective phase about $|\sigma_i\rangle$, including route-specific preparation and uncomputation. A stage uses phases about both endpoint projectors, giving cost
\begin{equation*}
O\!\left(L_i(C_i+C_{i+1})\right).
\end{equation*}

Since
\begin{equation*}
C_i+C_{i+1}
\leq
2\max\{C_i,C_{i+1}\},
\end{equation*}
substitution of Eq.~\eqref{eq:B14} gives the total transport cost
\begin{equation}
\label{eq:B24}
\widetilde O\!\left(
\sum_{i=0}^{\ell-1}
\frac{\max\{C_i,C_{i+1}\}}
{\sqrt{\underline q_i}}
\right).
\end{equation}

Equations~\eqref{eq:B21} and \eqref{eq:B24} prove Theorem 1.

\subsection{Overlap geometry and endpoint bias}

For the curved route, the selected sequence is the stationary-state sequence,
\begin{equation*}
|\sigma_i\rangle=|\pi_i\rangle.
\end{equation*}

For the flat route, the selected sequence consists of Perron states,
\begin{equation*}
|\sigma_i\rangle=|\mu_i\rangle.
\end{equation*}

Define their stationary-state biases by
\begin{equation}
\label{eq:B27}
b_i
=
D_{\mathrm{tr}}
\left(
|\mu_i\rangle\langle\mu_i|,
|\pi_i\rangle\langle\pi_i|
\right).
\end{equation}

By Eq.~\eqref{eq:A7},
\begin{equation*}
b_i
=
\sqrt{1-|\langle\mu_i|\pi_i\rangle|^2}.
\end{equation*}

Define
\begin{equation}
\label{eq:B31}
p_i
=
|\langle\pi_i|\pi_{i+1}\rangle|^2,
\qquad
q_i
=
|\langle\mu_i|\mu_{i+1}\rangle|^2.
\end{equation}

The Fubini--Study distance between normalized pure states is
\begin{equation*}
d_{\mathrm{FS}}(\psi,\varphi)
=\arccos|\langle\psi|\varphi\rangle|\in[0,\pi/2].
\end{equation*}

To see its triangle inequality, choose phases of $|u\rangle$ and $|w\rangle$ so that their overlaps with $|v\rangle$ are nonnegative, and write $x=d_{\mathrm{FS}}(u,v)$ and $y=d_{\mathrm{FS}}(v,w)$. Decomposing $|u\rangle$ and $|w\rangle$ along $|v\rangle$ and its orthogonal complement gives, by Cauchy--Schwarz,
\begin{equation*}
|\langle u|w\rangle|
\geq
\cos x\cos y-\sin x\sin y
=
\cos(x+y).
\end{equation*}

For $x+y\leq\pi/2$, this implies $d_{\mathrm{FS}}(u,w)\leq x+y$; for $x+y>\pi/2$ the inequality follows from the range of the distance.

By Eq.~\eqref{eq:A7}, $d_{\mathrm{FS}}(\mu_i,\pi_i)=\arcsin b_i$. Applying the triangle inequality along the stationary--Perron path gives
\begin{equation*}
\begin{aligned}
d_{\mathrm{FS}}(\mu_i,\mu_{i+1})
&\leq\arcsin b_i+\arccos\sqrt{p_i}\\
&\quad+\arcsin b_{i+1}.
\end{aligned}
\end{equation*}

Since $\cos^2$ decreases on $[0,\pi/2]$, it follows that
\begin{equation}
\label{eq:B35}
\begin{aligned}
q_i&\geq\cos^2\!\left(\min\{\pi/2,\theta_i\}\right),\\
\theta_i&=\arccos\sqrt{p_i}+\arcsin b_i+\arcsin b_{i+1}.
\end{aligned}
\end{equation}

For $x,y\in[0,\pi/2]$, $\cos(\min\{\pi/2,x+y\})\geq[\cos x-\sin y]_+$, where $[z]_+=\max\{z,0\}$. For $x+y\leq\pi/2$ this follows from $\cos x-\cos(x+y)=2\sin(x+y/2)\sin(y/2)\leq\sin y$; otherwise the right-hand side vanishes. Applying the inequality twice shows that Eq.~\eqref{eq:B35} is at least as strong as the linear certificate $q_i\geq[\sqrt{p_i}-b_i-b_{i+1}]_+^2$.

The angular certificate is sharp given only overlap and bias bounds. Let $\alpha=\arcsin b_i$, $\vartheta=\arccos\sqrt{p_i}$, and $\beta=\arcsin b_{i+1}$, and set $|v(t)\rangle=\cos t\,|0\rangle+\sin t\,|1\rangle$. If $\alpha+\vartheta+\beta\leq\pi/2$, the four states $|\mu_i\rangle$, $|\pi_i\rangle$, $|\pi_{i+1}\rangle$, and $|\mu_{i+1}\rangle$ can be taken at angles $0$, $\alpha$, $\alpha+\vartheta$, and $\alpha+\vartheta+\beta$, respectively. They attain equality in Eq.~\eqref{eq:B35}. If the total angle exceeds $\pi/2$ and the biases are upper bounds, the two bias angles can be reduced within their bounds so that the total angle is $\pi/2$, giving orthogonal selected states. The construction uses nonnegative real amplitudes, but does not establish optimality over states jointly realizable by Markov kernels satisfying the spectral filtering hypotheses.

Now suppose the forward-only transport starts from the stationary sample $|\pi_0\rangle$. The actual first selected target is $|\mu_0\rangle$, so
\begin{equation*}
\delta_0=b_0.
\end{equation*}

The transport theorem bounds the final error relative to $|\mu_\ell\rangle$. Applying the triangle inequality to the final stationary target,
\begin{equation*}
\begin{aligned}
D_{\mathrm{tr}}(\rho_\ell,\Pi_{\pi_\ell})
&\leq
D_{\mathrm{tr}}(\rho_\ell,\Pi_{\mu_\ell})
+
D_{\mathrm{tr}}(\Pi_{\mu_\ell},\Pi_{\pi_\ell})\\
&=
D_{\mathrm{tr}}(\rho_\ell,\Pi_{\mu_\ell})
+b_\ell.
\end{aligned}
\end{equation*}

Inserting Eq.~\eqref{eq:B21} and $\delta_0=b_0$ yields
\begin{equation}
\label{eq:B38}
D_{\mathrm{tr}}(\rho_\ell,\Pi_{\pi_\ell})
\leq
b_0+b_\ell+
\sum_{i=0}^{\ell-1}
\left[
2(L_i-1)\eta_i^{\mathrm{ph}}
+
\epsilon_i^{\mathrm{FP}}
\right].
\end{equation}

Equation~\eqref{eq:B38} proves the endpoint stationary-state accuracy bound of Sec.~\ref{sec:transport}.

\subsection{Route-specific preparation corollaries}

\textbf{Corollary 1 (Curved-route stationary-state preparation).}
Let $P_0,\ldots,P_\ell$ be ergodic nonreversible kernels with stationary QSAMPLEs $|\pi_i\rangle$. Assume that $|\pi_0\rangle$ is available, that coherent forward and time-reversed powered-transition oracles are supplied, and that the chosen powers $k_i$ satisfy the curved spectral-filtering assumptions. Let
\begin{equation}
\label{eq:B39}
0<\underline p_i
\leq
|\langle\pi_i|\pi_{i+1}\rangle|^2
\end{equation}
be certified overlap bounds.

With phase-cost factors
\begin{equation}
\label{eq:B40}
C_i^{\mathrm{curv}}
=
\frac{
c_{P_i^{k_i}}+
c_{(P_i^\star)^{k_i}}
}
{\sqrt{\gamma_{k_i}(P_i)}},
\end{equation}
the primitive-oracle complexity is
\begin{equation}
\label{eq:B41}
\widetilde O\!\left(
\sum_{i=0}^{\ell-1}
\frac{
\max\{C_i^{\mathrm{curv}},C_{i+1}^{\mathrm{curv}}\}
}
{\sqrt{\underline p_i}}
\right).
\end{equation}

The trace-distance error is the accumulated phase and FPAA error of Eq.~\eqref{eq:B21}, with $\delta_0=0$.

\textit{Proof.} Set $\sigma_i=\pi_i$ in Theorem 1. Then $q_i=p_i$, and the initial state gives zero initialization mismatch. The curved selective-phase construction supplies the costs in Eq.~\eqref{eq:B40}, with its logarithmic accuracy factors included in the $\widetilde O$ notation. Substitution into Eq.~\eqref{eq:B24} proves the cost bound. The error bound follows directly from Eq.~\eqref{eq:B21}. $\square$

\textbf{Corollary 2 (Flat-route approximation to the stationary QSAMPLE).}
Suppose each powered flat discriminant $D_{\mathrm{flat}}(P_i^{j_i})$ is primitive and satisfies all hypotheses of Proposition 2. Let $|\mu_i\rangle$ be its normalized Perron state, with certified overlaps
\begin{equation}
\label{eq:B42}
0<\underline q_i
\leq
|\langle\mu_i|\mu_{i+1}\rangle|^2.
\end{equation}

Define
\begin{equation}
\label{eq:B43}
C_i^{\mathrm{flat}}
=
\frac{
c_{P_i^{j_i}}
}
{\sqrt{\gamma_+(Q_{i,j_i})}}.
\end{equation}

The primitive forward-oracle complexity is
\begin{equation}
\label{eq:B44}
\widetilde O\!\left(
\sum_{i=0}^{\ell-1}
\frac{
\max\{C_i^{\mathrm{flat}},C_{i+1}^{\mathrm{flat}}\}
}
{\sqrt{\underline q_i}}
\right).
\end{equation}

Starting from $|\pi_0\rangle$, the final state satisfies Eq.~\eqref{eq:B38}.

\textit{Proof.} Apply Theorem 1 with $\sigma_i=\mu_i$. Proposition 2 supplies the phase costs in Eq.~\eqref{eq:B43}. Equations~\eqref{eq:B24} and \eqref{eq:B38} give the stated conclusions. $\square$

For a target stationary-state accuracy $\epsilon$, an error budget may reserve $\epsilon_{\mathrm{bias}}\geq b_0+b_\ell$ for the endpoint mismatch and distribute the remaining $\epsilon-\epsilon_{\mathrm{bias}}$ among the compiled phases and FPAA stages.

If an alternative initial state $\rho_0$ is supplied, the same theorem applies with
\begin{equation*}
\delta_0
=
D_{\mathrm{tr}}
\left(
\rho_0,
|\mu_0\rangle\langle\mu_0|
\right).
\end{equation*}

The sufficient spectral conditions for the positive overlap bounds $\underline q_i$ are developed in Appendix~\ref{app:certificates}.

Theorem 1 also applies to mixed selectors: a known initial state $|\chi\rangle$ with an available phase, flat interior targets, and a curved stationary final target. Use their actual adjacent overlap bounds and route-specific phase costs in Eq.~\eqref{eq:B24}. The error is Eq.~\eqref{eq:B21} with $\delta_0=D_{\mathrm{tr}}(\rho_0,|\chi\rangle\langle\chi|)$ and no endpoint Perron bias. The final selector requires clean powered reverse access.

%% file: appendix_C.tex
\section{Selective phases from filter responses}
\label{app:phase-response}
We prove Proposition 1 (Sec.~\ref{subsec:phase-interface}) and its extension to complete success projectors. Ancilla-efficient approximate reflections about unitary eigenspaces have been studied in Refs.~\cite{ChowdhurySubasiSomma2018,ClaudonReflect2025}. In the selective-phase compiler of Ref.~\cite{Zhao2026} (Proposition III.4), the target is accepted exactly and the filter response commutes with the target projector. The identity below also applies when these properties do not hold.

\subsection{Filter--phase--uncompute operation}

Let $C$ be a unitary acting on a signal qubit and a data Hilbert space $\mathcal H$. Write its first block column as
\begin{equation}
\label{eq:C1}
C(|0\rangle\otimes I)
=
|0\rangle\otimes A
+
|1\rangle\otimes B,
\end{equation}
with
\begin{equation*}
A^\dagger A+B^\dagger B=I.
\end{equation*}

The accepted filter-response operator is
\begin{equation}
\label{eq:C3}
G=A^\dagger A,
\qquad 0\leq G\leq I.
\end{equation}

\begin{figure}[t]
\centering
\includegraphics[width=\columnwidth]{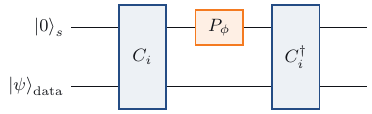}
\caption{Selective-phase interface. A filtering unitary $C_i$ acts on the clean signal qubit and data register, the phase $P_\phi=e^{i\phi}|0\rangle\langle0|+|1\rangle\langle1|$ is applied to the signal, and $C_i^\dagger$ uncomputes the filter. On clean auxiliary inputs the net action approximates $S_\phi(\Lambda)$ on the target projector. Route-specific oracle preparation and uncomputation are accounted for separately.}
\label{fig:interface}
\end{figure}

Apply a phase $e^{i\phi}$ to the successful signal state and uncompute the filter, as shown in Fig.~\ref{fig:interface}:
\begin{equation}
\label{eq:C4}
U_\phi
=
C^\dagger(P_\phi\otimes I)C,
\end{equation}
where
\begin{equation*}
P_\phi
=
e^{i\phi}|0\rangle\langle0|
+
|1\rangle\langle1|.
\end{equation*}

For a target projector $\Lambda$, the ideal selective phase is
\begin{equation*}
S_\phi(\Lambda)
=
I+(e^{i\phi}-1)\Lambda.
\end{equation*}

We compare the implemented and ideal operations on inputs with a clean signal register:
\begin{equation}
\label{eq:C7}
E_\phi
=
U_\phi(|0\rangle\otimes I)
-
|0\rangle\otimes S_\phi(\Lambda).
\end{equation}

\subsection{Proof of Proposition 1}

Let $J:\mathcal H\longrightarrow\mathcal K$ be an input isometry, $C$ a filtering unitary on $\mathcal K$, and $\Pi_{\mathrm{succ}}$ an orthogonal projector onto the successful output subspace. Define
\begin{equation*}
\begin{aligned}
G&=J^\dagger C^\dagger\Pi_{\mathrm{succ}}CJ,\\
U_\phi&=C^\dagger(I+z\Pi_{\mathrm{succ}})C,
\end{aligned}
\end{equation*}
where $z=e^{i\phi}-1$. Since $J$ is an isometry, $0\leq G\leq I$.

For an orthogonal target projector $\Lambda$ with $\Lambda^\perp=I-\Lambda$, the error operator is
\begin{equation*}
\begin{aligned}
E_\phi
&=U_\phi J-JS_\phi(\Lambda)\\
&=zC^\dagger(\Pi_{\mathrm{succ}}CJ-CJ\Lambda)\\
&=zC^\dagger\big[
\Pi_{\mathrm{succ}}CJ\Lambda^\perp
\\
&\qquad\qquad
-(I-\Pi_{\mathrm{succ}})CJ\Lambda
\big].
\end{aligned}
\end{equation*}

The two terms inside the brackets have orthogonal output supports. By unitarity of $C$ and the isometry condition $J^\dagger J=I$,
\begin{equation}
\label{eq:C16}
\begin{aligned}
E_\phi^\dagger E_\phi
=|z|^2\big[
&\Lambda^\perp G\Lambda^\perp\\
&+\Lambda(I-G)\Lambda
\big].
\end{aligned}
\end{equation}

Both terms are positive semidefinite and act on orthogonal subspaces. With $\delta_{\mathrm{tar}}=\|\Lambda(I-G)\Lambda\|$ and $\delta_{\mathrm{bulk}}=\|\Lambda^\perp G\Lambda^\perp\|$, we obtain
\begin{equation}
\label{eq:C20}
\|E_\phi\|
=
|e^{i\phi}-1|
\sqrt{\max\{\delta_{\mathrm{tar}},\delta_{\mathrm{bulk}}\}}.
\end{equation}

Taking $J=|0\rangle\otimes I$ and $\Pi_{\mathrm{succ}}=|0\rangle\langle0|\otimes I$ gives $G=A^\dagger A$ and proves Proposition 1. $\square$

The quantities $\delta_{\mathrm{tar}}$ and $\delta_{\mathrm{bulk}}$ are the worst-case target rejection and bulk acceptance probabilities, respectively. For fixed $C$, $J$, and $\Pi_{\mathrm{succ}}$, uniform phase accuracy $\sup_{\phi\in\mathbb R}\|E_\phi\|\leq\eta$ is equivalent to $\max\{\delta_{\mathrm{tar}},\delta_{\mathrm{bulk}}\}\leq\eta^2/4$.

\subsection{Complete success projectors}

The same argument applies when success is specified by a projector on several auxiliary registers rather than by a single signal-qubit outcome.

Let
\begin{equation*}
J_{\mathrm{in}},J_{\mathrm{out}}:
\mathcal H\longrightarrow\mathcal K
\end{equation*}
be isometries identifying the input and successful output embeddings of the same system space, and let $C$ be the filtering unitary on $\mathcal K$. The complete successful output subspace is
\begin{equation*}
\Pi_{\mathrm{succ}}
=
J_{\mathrm{out}}J_{\mathrm{out}}^\dagger.
\end{equation*}

The accepted operator is
\begin{equation*}
F
=
J_{\mathrm{out}}^\dagger C J_{\mathrm{in}}.
\end{equation*}

Its positive response on the input system is therefore
\begin{equation}
\label{eq:C23}
G
=
F^\dagger F
=
J_{\mathrm{in}}^\dagger
C^\dagger\Pi_{\mathrm{succ}}C
J_{\mathrm{in}}.
\end{equation}

Applying the phase to the complete successful output subspace and uncomputing gives
\begin{equation*}
U_\phi
=
C^\dagger
\left[
I+(e^{i\phi}-1)\Pi_{\mathrm{succ}}
\right]
C.
\end{equation*}

The exact error identity in Eq.~\eqref{eq:C20} applies with $J=J_{\mathrm{in}}$ and the response operator in Eq.~\eqref{eq:C23}, without requiring commutation between $G$ and $\Lambda$.

The input and output embeddings need not coincide. The phase is applied to the physical output-success projector $\Pi_{\mathrm{succ}}$, and the error bound includes any leakage outside the input embedding after uncomputation.

%% file: appendix_D.tex
\section{Flat-discriminant spectral construction}
\label{app:flat}
We prove Proposition 2 (Sec.~\ref{subsec:flat-route}) and the associated stationary-state bias estimate.

\subsection{Flat discriminants and leading-state separation}

For a positive integer $j$, define
\begin{equation}
\label{eq:D1}
D_j
=
D(P^j,P^j),
\end{equation}
with entries
\begin{equation}
\label{eq:D2}
D_j(x,y)
=
\sqrt{P^j(x,y)P^j(y,x)}.
\end{equation}

The matrix $D_j$ is real, symmetric, and entrywise nonnegative. Since it is a compression of the swap unitary,
\begin{equation*}
\|D_j\|\leq1.
\end{equation*}

Suppose $D_j$ is primitive. Perron--Frobenius theory gives a unique normalized strictly positive leading eigenvector $|\mu_j\rangle$, with
\begin{equation}
\label{eq:D4}
D_j|\mu_j\rangle
=
\omega_j|\mu_j\rangle,
\qquad
0<\omega_j\leq1.
\end{equation}

Write $\mu_j(x)=\langle x|\mu_j\rangle>0$, and define
\begin{equation*}
\nu_j(x)=\mu_j(x)^2.
\end{equation*}

The geometric reversibilization is
\begin{equation}
\label{eq:D6}
Q_j(x,y)
=
\frac{D_j(x,y)\mu_j(y)}
{\omega_j\mu_j(x)}.
\end{equation}

Using Eq.~\eqref{eq:D4},
\begin{equation*}
\sum_yQ_j(x,y)
=
\frac{(D_j\mu_j)(x)}
{\omega_j\mu_j(x)}
=1,
\end{equation*}
so $Q_j$ is stochastic.

Furthermore,
\begin{equation*}
\begin{aligned}
\nu_j(x)Q_j(x,y)
&=
\frac{\mu_j(x)\mu_j(y)D_j(x,y)}{\omega_j}\\
&=
\nu_j(y)Q_j(y,x),
\end{aligned}
\end{equation*}
where symmetry of $D_j$ was used. Thus $Q_j$ is reversible with stationary distribution $\nu_j$.

Let $M_{\mu_j}=\operatorname{diag}(\mu_j(x))$. Then
\begin{equation}
\label{eq:D9}
Q_j
=
\frac{1}{\omega_j}
M_{\mu_j}^{-1}D_jM_{\mu_j}.
\end{equation}

Hence $Q_j$ and $D_j/\omega_j$ are similar, and
\begin{equation*}
\operatorname{spec}(D_j)
=
\omega_j\operatorname{spec}(Q_j).
\end{equation*}

Writing $\lambda_2$ for the second-largest eigenvalue in algebraic order, the geometric Poincar\'e gap is
\begin{equation}
\label{eq:D11}
\gamma_+(Q_j)
=
1-\lambda_2(Q_j)
=
1-\frac{\lambda_2(D_j)}{\omega_j}.
\end{equation}

To implement the leading-state selection principle of Ref.~\cite{Claudon2025}, we construct a bounded polynomial separating the target and bulk spectral intervals.

\textbf{Lemma 3 (Bounded interval filter).}
For $0<\delta<2$ and $0<\eta_f<1$, there is a real polynomial $r$ such that
\begin{equation}
\label{eq:D-interval-filter}
\begin{aligned}
0\leq r(x)&\leq1 &&(-1\leq x\leq1),\\
r(x)&\leq\eta_f &&(-1\leq x\leq1-\delta),\\
r(x)&\geq1-\eta_f &&(1-\delta/16\leq x\leq1).
\end{aligned}
\end{equation}

Its degree is $2mh$, where $m=\lceil\delta^{-1/2}\rceil$ and $h$ is any positive odd integer with $h\geq8\ln(1/\eta_f)$. In particular, the degree is $O(\delta^{-1/2}[1+\ln(1/\eta_f)])$.

To construct it, let $T_m$ be the Chebyshev polynomial of the first kind and set
\begin{equation}
\label{eq:D-filter-construction}
\begin{aligned}
t(x)&=\frac{2x+\delta}{2-\delta},\qquad
u(x)=\left[\frac{T_m(t(x))}{T_m(t(1))}\right]^2,\\
r(x)&=B_h(u(x)),\\
B_h(u)&=\sum_{k=(h+1)/2}^{h}\binom hk u^k(1-u)^{h-k}.
\end{aligned}
\end{equation}

On the rejection interval, $t(x)\in[-1,1]$. Since
$\operatorname{arcosh}t(1)=2\operatorname{artanh}\sqrt{\delta/2}\geq\sqrt{2\delta}$,
we have $T_m(t(1))\geq\cosh\sqrt2>2$, so $u\leq1/4$ there. On all of $[-1,1]$, $0\leq u\leq1$. On the acceptance interval, $t(x)>1$ and $u'$ is nonnegative and increasing, while
\begin{equation*}
u'(1)\leq\frac{\sqrt2m}{\sqrt\delta}
\leq\frac{2+\sqrt2}{\delta}<\frac4\delta.
\end{equation*}

Thus $u(x)\geq3/4$ whenever $x\geq1-\delta/16$. For $u\leq1/4$, the binomial tail bound~\cite{Hoeffding1963} gives $B_h(u)\leq e^{-h/8}$; symmetry gives $1-B_h(u)\leq e^{-h/8}$ for $u\geq3/4$. This proves the lemma. $\square$

The construction needs only the interval parameter $\delta$ and the requested accuracy, not the target eigenvector or exact Perron eigenvalue. More sharply, one may choose $h$ by evaluating the two binomial tails at certified bounds on the bulk maximum and target minimum of $u$; Appendix~\ref{app:verification} uses this constructive choice.

For the discriminant, put
\begin{equation*}
\delta_j=1-\lambda_2(D_j)
=1-\omega_j+\omega_j\gamma_+(Q_j).
\end{equation*}

The acceptance condition $1-\omega_j\leq c_\star\delta_j$, with $c_\star=1/16$, is equivalent to
\begin{equation}
\label{eq:D12}
\omega_j
\geq
\frac{1-c_\star}
{1-c_\star[1-\gamma_+(Q_j)]}.
\end{equation}

It places the Perron eigenvalue in the acceptance interval and the remaining spectrum in the rejection interval. Consequently,
\begin{equation}
\label{eq:D13}
|r_j(x)|\leq1\qquad(x\in[-1,1]),
\end{equation}
\begin{equation}
\label{eq:D14}
r_j(\omega_j)\geq1-\eta_f,
\end{equation}
and
\begin{equation}
\label{eq:D15}
|r_j(\lambda)|\leq\eta_f
\quad\text{for every non-target eigenvalue }\lambda.
\end{equation}

Since $\delta_j<2$, the premise implies $\omega_j>7/8$ and $\delta_j\geq\omega_j\gamma_+(Q_j)$. Hence
\begin{equation}
\label{eq:D16}
d_j=O\!\left(
\frac{1+\ln(1/\eta_f)}{\sqrt{\gamma_+(Q_j)}}
\right).
\end{equation}

Operationally, certified bounds $\lambda_2(D_j)\leq a_j<1$ and $\omega_j\geq w_j$, with $a_j>-1$ and $1-w_j\leq(1-a_j)/16$, suffice: set $\delta=1-a_j$ in Lemma 3. The resulting degree is controlled by the supplied separation $1-a_j$, rather than an unavailable exact gap. The case $a_j\leq0$ is included without shrinking $\delta$: for $1\leq\delta<2$, $m=1$ and $u(x)=[(2x+\delta)/(2+\delta)]^2$. This retains the acceptance interval required by the Perron enclosure.

\subsection{Conjugate eigenphases and projector compression}

Let $\square_j=\square_{P^j}$. Since
\begin{equation*}
D_j=\square_j^\dagger S\square_j,
\end{equation*}

the associated Szegedy walk is
\begin{equation}
\label{eq:D19}
W_j
=
(2\square_j\square_j^\dagger-I)S.
\end{equation}

Define
\begin{equation*}
\mathcal K_j
=
\operatorname{Ran}(\square_j)
+
\operatorname{Ran}(S\square_j).
\end{equation*}

The subspace $\mathcal K_j$ is invariant under $W_j$. The embedded states $\square_j|\psi\rangle$ lie in this subspace, so the spectral action relevant to the selector is its restriction to $\mathcal K_j$.

Let $|v\rangle$ be a normalized eigenvector of $D_j$ with
\begin{equation}
\label{eq:D21}
D_j|v\rangle
=
\lambda|v\rangle,
\qquad
\lambda=\cos\theta,
\qquad
|\lambda|<1.
\end{equation}

Set
\begin{equation*}
|u\rangle=\square_j|v\rangle,
\qquad
|s\rangle=S\square_j|v\rangle.
\end{equation*}

The encoding identity gives
\begin{equation*}
\langle u|s\rangle
=
\langle v|D_j|v\rangle
=
\cos\theta.
\end{equation*}

The walk acts on these vectors as
\begin{equation*}
W_j|s\rangle=|u\rangle,
\end{equation*}
and
\begin{equation*}
W_j|u\rangle
=
2\cos\theta\,|u\rangle-|s\rangle.
\end{equation*}

The two normalized eigenvectors in this invariant plane are
\begin{equation}
\label{eq:D26}
|w_\pm(v)\rangle
=
\frac{
(e^{\pm i\theta}-S)\square_j|v\rangle
}{
\sqrt2\sin\theta
},
\end{equation}
with
\begin{equation}
\label{eq:D27}
W_j|w_\pm(v)\rangle
=
e^{\pm i\theta}|w_\pm(v)\rangle.
\end{equation}

Subtracting the two expressions in Eq.~\eqref{eq:D26} gives
\begin{equation*}
|w_+(v)\rangle-|w_-(v)\rangle
=
i\sqrt2\,\square_j|v\rangle,
\end{equation*}
so
\begin{equation}
\label{eq:D29}
\square_j|v\rangle
=
\frac{
|w_+(v)\rangle-|w_-(v)\rangle
}{
i\sqrt2
}.
\end{equation}

Let $\Lambda_v$ be the projector onto the span of this conjugate eigenphase pair,
\begin{equation}
\label{eq:D30}
\Lambda_v
=
|w_+(v)\rangle\langle w_+(v)|
+
|w_-(v)\rangle\langle w_-(v)|.
\end{equation}

Orthogonality of distinct discriminant eigenspaces and Eq.~\eqref{eq:D29} imply
\begin{equation*}
\Lambda_v\square_j|v\rangle
=
\square_j|v\rangle,
\end{equation*}
while $\Lambda_v\square_j|v'\rangle=0$ for normalized discriminant eigenvectors orthogonal to $v$ in the other invariant planes.

Compression onto the original state space therefore gives
\begin{equation}
\label{eq:D32}
\square_j^\dagger\Lambda_v\square_j
=
|v\rangle\langle v|.
\end{equation}

For the leading Perron state, the conjugate eigenphases are $\pm\theta_j$, with $\theta_j=\arccos\omega_j$.

The complete pair is the target subspace. The selective phase depends on the positive response on this subspace; the accepted amplitude need not give the two eigenphases the same complex phase. 

When $\omega_j=1$, the conjugate phases coalesce at zero. In this case
\begin{equation*}
\langle\square_j\mu_j|
S\square_j\mu_j\rangle
=
\langle\mu_j|D_j|\mu_j\rangle
=
1.
\end{equation*}

Both $\square_j|\mu_j\rangle$ and $S\square_j|\mu_j\rangle$ are normalized, so equality in the unit-overlap condition implies
\begin{equation*}
S\square_j|\mu_j\rangle
=
\square_j|\mu_j\rangle.
\end{equation*}

Since $\square_j|\mu_j\rangle$ lies in $\operatorname{Ran}(\square_j)$,
\begin{equation*}
W_j\square_j|\mu_j\rangle
=
(2\square_j\square_j^\dagger-I)
S\square_j|\mu_j\rangle
=
\square_j|\mu_j\rangle.
\end{equation*}

Thus the embedded Perron state is a $+1$ eigenstate of $W_j$. Under primitivity, the Perron eigenvalue of $D_j$ is simple; hence the corresponding $+1$ eigenspace in the invariant walk sector is one-dimensional. If $\Pi_{W_j=1}$ denotes its projector, then
\begin{equation}
\label{eq:D33d}
\square_j^\dagger
\Pi_{W_j=1}
\square_j
=
|\mu_j\rangle\langle\mu_j|.
\end{equation}

\subsection{Bounded-Laurent transformation and GQSP realization}

The real polynomial $r_j$ is bounded on $[-1,1]$. To evaluate it through the unitary walk $W_j$, define
\begin{equation}
\label{eq:D34}
R_j(z)
=
r_j\!\left(\frac{z+z^{-1}}2\right).
\end{equation}

For $z=e^{i\vartheta}$,
\begin{equation*}
R_j(e^{i\vartheta})
=
r_j(\cos\vartheta).
\end{equation*}

Therefore, Eq.~\eqref{eq:D13} implies
\begin{equation}
\label{eq:D36}
\sup_{|z|=1}|R_j(z)|\leq1.
\end{equation}

Since $r_j$ has degree $d_j$, the Laurent polynomial $R_j$ contains powers between $-d_j$ and $d_j$. Define the ordinary polynomial
\begin{equation}
\label{eq:D37}
\mathcal P_j(z)
=
z^{d_j}R_j(z).
\end{equation}

It has degree at most $2d_j$, and multiplication by $z^{d_j}$ preserves its modulus on the unit circle. Thus
\begin{equation}
\label{eq:D38}
\sup_{|z|=1}|\mathcal P_j(z)|
\leq1.
\end{equation}

The GQSP existence theorem of Motlagh and Wiebe \cite{MotlaghWiebe2024} applies to this bounded polynomial. Equivalently, the nonnegative trigonometric polynomial $1-|\mathcal P_j(z)|^2$ admits a complementary-polynomial factorization on the unit circle \cite{BerntsonSunderhauf2025}, providing a unitary one-signal-qubit GQSP completion.

Consequently, there is a GQSP unitary $C_j^{\mathrm{GQSP}}$ whose upper-left signal block is $\mathcal P_j(W_j)$.

Define
\begin{equation*}
X_j
=
\frac{W_j+W_j^\dagger}{2}.
\end{equation*}

Functional calculus gives
\begin{equation*}
R_j(W_j)
=
r_j(X_j).
\end{equation*}

The undelayed GQSP unitary $C_j^{\mathrm{GQSP}}$ has accepted amplitude operator
\begin{equation}
\label{eq:D43}
A_j=\mathcal P_j(W_j)=W_j^{d_j}r_j(X_j).
\end{equation}

Its positive response is $A_j^\dagger A_j=r_j(X_j)^2$. Thus the two conjugate eigenphases may receive different amplitude phases $e^{\pm id_j\theta}$, but have the same positive response $r_j(\cos\theta)^2$.

An unconditional delay would define
$C_j=(I_{\mathrm{sig}}\otimes W_j^{-d_j})C_j^{\mathrm{GQSP}}$.
It commutes with the signal phase and cancels exactly:
\begin{equation}
\label{eq:D-delay-cancellation}
\begin{aligned}
C_j^\dagger(P_\phi\otimes I)C_j
&=(C_j^{\mathrm{GQSP}})^\dagger
(P_\phi\otimes I)C_j^{\mathrm{GQSP}}.
\end{aligned}
\end{equation}

We therefore use the undelayed circuit. A degree-$2d_j$ GQSP sequence and its inverse require at most
\begin{equation}
\label{eq:D44}
4d_j
\end{equation}
controlled applications of $W_j$ or $W_j^\dagger$ per selective phase.

For comparison, take the same real degree-$d$ polynomial $r$, walk oracle, and register conventions. The GQET core of Ref.~\cite{Claudon2025} encodes $r(D)/\alpha_r$ in the successful signal and encoded subspace, using one signal qubit and $d$ controlled walk calls ($2d$ for filter and inverse). The Laurent core encodes $W^d r(X)$ with signal success and an outer walk embedding, using one signal qubit and at most $2d$ controlled walk calls ($4d$ for filter and inverse). Its ideal bounded response needs no extra normalization, whereas $\alpha_r$ is polynomial-dependent and may equal one. These are core counts; success-projector operations, workspace, and any normalization correction must be counted separately.

If $\alpha_r>1$ and $|r|\leq1$, using the GQET amplitude directly in filter--phase--uncompute with its complete success projector gives target acceptance at most $\alpha_r^{-2}$. By Proposition 1,
\begin{equation*}
\|E_\phi\|\geq |e^{i\phi}-1|\sqrt{1-\alpha_r^{-2}}.
\end{equation*}
This explains the importance of normalization for this use of a filter, but does not rule out other GQET reflection constructions.

\subsection{Finite-precision phase synthesis}

The exact GQSP construction accepts the unit-circle bound in Eq.~\eqref{eq:D38}, including equality. For finite-precision coefficient and rotation synthesis \cite{BerntsonSunderhauf2025}, introduce an optional interior margin
\begin{equation*}
\widetilde r_j=\rho_j r_j,
\qquad
\rho_j=1-\kappa_j,
\qquad
0\leq\kappa_j<1.
\end{equation*}

The same Laurent construction applies to $\widetilde r_j$. Its accepted block is
\begin{equation*}
A_j=\rho_jW_j^{d_j}r_j(X_j).
\end{equation*}

Because $X_j$ is Hermitian and $r_j$ is real,
\begin{equation}
\label{eq:D47}
G_j=A_j^\dagger A_j
=
\rho_j^2r_j(X_j)^2.
\end{equation}

On the Perron eigenphase pair, Eq.~\eqref{eq:D14} gives
\begin{equation}
\label{eq:D48}
q_{\mathrm{tar}}
\geq
\rho_j^2(1-\eta_f)^2.
\end{equation}

On the remaining relevant walk spectrum,
\begin{equation}
\label{eq:D49}
q_{\mathrm{bulk}}
\leq
\rho_j^2\eta_f^2.
\end{equation}

For $\eta_f\leq1/2$, Proposition 1 gives
\begin{equation*}
\eta_{\phi,j}^{\mathrm{filt}}
\leq
2\sqrt{
1-\rho_j^2(1-\eta_f)^2
}.
\end{equation*}

Since $\rho_j=1-\kappa_j$,
\begin{equation*}
\begin{aligned}
1-\rho_j^2(1-\eta_f)^2
&\leq
1-(1-\kappa_j)^2(1-\eta_f)^2\\
&\leq
2(\kappa_j+\eta_f).
\end{aligned}
\end{equation*}

Therefore,
\begin{equation}
\label{eq:D52}
\eta_{\phi,j}^{\mathrm{filt}}
\leq
2\sqrt{2(\kappa_j+\eta_f)}.
\end{equation}

A desired filter-induced phase error $\eta_\phi$ is obtained, for example, by choosing
\begin{equation}
\label{eq:D53}
\kappa_j,\eta_f=O(\eta_\phi^2).
\end{equation}

Coefficient discretization, rotation synthesis, and oracle errors are included in the full phase-error budget. Their accumulation is treated in Appendix~\ref{app:resources}.

The margin does not change the asymptotic polynomial degree. At $\kappa_j=0$, the exact bounded filter is recovered.

\subsection{Flat selective phase and stationary-state bias}

Let $\Lambda_{\mu_j}$ project onto the Perron conjugate eigenphase pair, and define the clean embedding
\begin{equation*}
J_j
=
|0\rangle_{\mathrm{sig}}\otimes\square_j.
\end{equation*}

Then
\begin{equation*}
J_j^\dagger J_j=I.
\end{equation*}

By Eq.~\eqref{eq:D32},
\begin{equation}
\label{eq:D56}
\square_j^\dagger
\Lambda_{\mu_j}
\square_j
=
|\mu_j\rangle\langle\mu_j|.
\end{equation}

The phase circuit obtained from the filter--phase--uncompute construction acts on the target pair with the response in Eq.~\eqref{eq:D48} and on the remaining relevant spectrum with the response in Eq.~\eqref{eq:D49}.

Proposition 1 therefore gives
\begin{equation}
\label{eq:D57}
\left\|
\widetilde S_{\phi,j}^{\mathrm{flat}}J_j
-
J_jS_\phi(|\mu_j\rangle\langle\mu_j|)
\right\|
\leq\eta_{\phi,j},
\end{equation}
where $\eta_{\phi,j}$ includes the filter, coefficient, rotation, and oracle contributions.

By Eqs.~\eqref{eq:D16}, \eqref{eq:D44}, and \eqref{eq:D53}, the implementation uses one GQSP signal qubit and
\begin{equation*}
O\!\left(
\frac{1}{\sqrt{\gamma_+(Q_j)}}
\log\frac1{\eta_\phi}
\right)
\end{equation*}
controlled walk applications. The one-signal-qubit count applies to this GQSP compiler; it excludes the two transition registers and oracle workspace.

To obtain the fixed clean-register interface, conjugate the walk-space phase by $O_{P^j}$ and its inverse. These two outer operations prepare $\square_j$ from the zero transition register and return to the common embedding $J$. Each walk application itself uses one $O_{P^j}$ and one $O_{P^j}^\dagger$. Thus one selective phase uses at most $8d_j+2$ clean powered-transition oracle calls, including preparation and uncomputation. The primitive-oracle cost is
\begin{equation}
\label{eq:D59}
O\!\left(
\frac{c_{P^j}}{\sqrt{\gamma_+(Q_j)}}
\log\frac1{\eta_\phi}
\right).
\end{equation}

If a specified clean endpoint construction establishes $c_{P^j}=O(jc_P)$, this becomes
\begin{equation*}
\widetilde O\!\left(
\frac{jc_P}{\sqrt{\gamma_+(Q_j)}}
\right),
\end{equation*}
with the accuracy logarithms suppressed. This completes Proposition 2.

Expand the stationary QSAMPLE $|\pi\rangle$ in an orthonormal eigenbasis of the symmetric matrix $D_j$. Let $|\mu_j\rangle$ be the leading eigenvector, and write
\begin{equation*}
|\pi\rangle
=
a|\mu_j\rangle
+
\sqrt{1-|a|^2}\,|\mu_j^\perp\rangle,
\end{equation*}
where $|\mu_j^\perp\rangle$ is normalized and orthogonal to $|\mu_j\rangle$, and $a=\langle\mu_j|\pi\rangle$. If $|a|=1$, the bias is zero and the bound below is immediate.

Since the largest eigenvalue on the orthogonal complement is $\lambda_2(D_j)$,
\begin{equation*}
\begin{aligned}
\langle\pi|D_j|\pi\rangle
&\leq
\omega_j|a|^2
+
\lambda_2(D_j)(1-|a|^2)\\
&=
\omega_j
-
[\omega_j-\lambda_2(D_j)](1-|a|^2).
\end{aligned}
\end{equation*}

Using
\begin{equation*}
\omega_j-\lambda_2(D_j)
=
\omega_j\gamma_+(Q_j),
\end{equation*}
we obtain
\begin{equation}
\label{eq:D64}
1-|\langle\pi|\mu_j\rangle|^2
\leq
\frac{
\omega_j-\langle\pi|D_j|\pi\rangle
}{
\omega_j\gamma_+(Q_j)
}.
\end{equation}

Define
\begin{equation}
\label{eq:D65}
\Delta_j
=
1-\langle\pi|D_j|\pi\rangle.
\end{equation}

Since $D_j$ is entrywise nonnegative and $|\pi\rangle$ has nonnegative amplitudes,
\begin{equation*}
0\leq1-\Delta_j
=
\langle\pi|D_j|\pi\rangle\leq\omega_j\leq1.
\end{equation*}

As $\omega_j>0$, this gives
\begin{equation*}
\frac{\omega_j-(1-\Delta_j)}{\omega_j}
=
1-\frac{1-\Delta_j}{\omega_j}\leq\Delta_j.
\end{equation*}

The pure-state trace-distance bias satisfies
\begin{equation}
\label{eq:D69}
b_j^2
=
1-|\langle\pi|\mu_j\rangle|^2.
\end{equation}

The spectral decomposition also gives, with $n_j=\omega_j-\langle\pi|D_j|\pi\rangle$ and orthonormal eigenvectors $v_k$,
\begin{equation*}
n_j=\sum_{k>1}(\omega_j-\lambda_k)|\langle v_k|\pi\rangle|^2.
\end{equation*}
Since the weights sum to $b_j^2$, we have $n_j\leq[\omega_j-\lambda_{\min}(D_j)]b_j^2\leq2b_j^2$. This proves the lower bias bound in Eq.~\eqref{eq:flat-bias-bound}. Combining Eq.~\eqref{eq:D64} with the deficit inequality yields
\begin{equation}
\label{eq:D70}
b_j^2
\leq
\frac{\Delta_j}
{\gamma_+(Q_j)}.
\end{equation}

%% file: appendix_E.tex
\section{Curved-discriminant spectral construction}
\label{app:curved}
We prove the curved selective-phase construction and its resource bound from Sec.~\ref{subsec:curved-route}.

\subsection{Curved discriminant and stationary singular target}

Let $P$ be ergodic with full-support stationary distribution $\pi$. Write
\begin{equation*}
M_\pi=\operatorname{diag}(\pi(x)).
\end{equation*}

The curved discriminant is
\begin{equation*}
D_{\mathrm{curv}}
=
D(P,P^\star).
\end{equation*}

Using Eq.~\eqref{eq:A19},
\begin{equation*}
\begin{aligned}
D_{\mathrm{curv}}(x,y)
&=
\sqrt{
P(x,y)\frac{\pi(x)P(x,y)}{\pi(y)}
}\\
&=
\sqrt{\frac{\pi(x)}{\pi(y)}}P(x,y).
\end{aligned}
\end{equation*}

Therefore,
\begin{equation}
\label{eq:E4}
D_{\mathrm{curv}}
=
M_\pi^{1/2}PM_\pi^{-1/2}.
\end{equation}

Since $P$ is row stochastic, $P\mathbf1=\mathbf1$. Noting that $M_\pi^{-1/2}|\pi\rangle=\mathbf1$, we obtain
\begin{equation}
\label{eq:E7}
D_{\mathrm{curv}}|\pi\rangle
=
|\pi\rangle.
\end{equation}

Similarly, stationarity gives
\begin{equation}
\label{eq:E8}
\langle\pi|D_{\mathrm{curv}}
=
\langle\pi|.
\end{equation}

Thus $|\pi\rangle$ is both a right and left singular vector with singular value one.

The curved discriminant is a contraction. To see this, let $f$ be a complex function on $\Omega$. Jensen's inequality and stationarity give
\begin{equation*}
\begin{aligned}
\|Pf\|_{L^2(\pi)}^2
&=
\sum_x\pi(x)
\left|\sum_yP(x,y)f(y)\right|^2\\
&\leq
\sum_x\pi(x)\sum_yP(x,y)|f(y)|^2\\
&=
\sum_y\pi(y)|f(y)|^2.
\end{aligned}
\end{equation*}

Thus
\begin{equation*}
\|D_{\mathrm{curv}}\|\leq1.
\end{equation*}

For a positive integer $k$, the powered curved discriminant is
\begin{equation}
\label{eq:E11}
A_k
=
D(P^k,(P^\star)^k)
=
M_\pi^{1/2}P^kM_\pi^{-1/2}.
\end{equation}

Its singular-value-one target remains $|\pi\rangle$.

\subsection{Powered singular-value filtering}

Define the $k$-step multiplicative-reversibilization gap
\begin{equation}
\label{eq:E12}
\gamma_k(P)
=
\gamma\!\left(P^k(P^\star)^k\right),
\end{equation}
where $\gamma$ denotes the absolute spectral gap of the reversible kernel in the argument.

Using the weighted similarity transformation,
\begin{equation*}
A_kA_k^\dagger
=
M_\pi^{1/2}
P^k(P^\star)^k
M_\pi^{-1/2}.
\end{equation*}

The eigenvalues of $A_kA_k^\dagger$ are the squared singular values of $A_k$. In particular, if the stationary singular value is isolated and the next singular value is $s_2(A_k)$, then
\begin{equation}
\label{eq:E14}
s_2(A_k)^2
=
1-\gamma_k(P).
\end{equation}

We invoke the fast-forwarding singular-value filter of Claudon, Piquemal, and Monmarch\'e \cite{Claudon2025} under the assumptions used here: the stationary singular value $1$ is isolated, the powered multiplicative-reversibilization gap satisfies $\gamma_k(P)>0$ with a certified lower bound, and coherent clean access to the forward and time-reversed powered-transition encodings is available. The filter supplies a block $F_k$ satisfying
\begin{equation}
\label{eq:E15}
F_k|\pi\rangle=|\pi\rangle,
\end{equation}
and
\begin{equation}
\label{eq:E16}
\|F_k-\Pi_\pi\|\leq\eta_f
\end{equation}
on the relevant right-singular space.

Its polynomial degree obeys
\begin{equation}
\label{eq:E17}
d_k
=
O\!\left(
\frac1{\sqrt{\gamma_k(P)}}
\log\frac1{\eta_f}
\right).
\end{equation}

Polynomial spectral transformations provide a general framework for such filters \cite{Gilyen2019,Martyn2021}; related mathematical results on polynomial approximation of the sign function are given in Ref.~\cite{EremenkoYuditskii2007}.

The filtering theorem supplies a real fixed-parity polynomial $\upsilon_k$ bounded in magnitude by one on all of $[-1,1]$, with unit magnitude at the stationary singular value and magnitude at most $\eta_f$ on the remaining singular spectrum. This global bound is the input required for the projected polynomial transformation.

The following Hermitianization makes the target support of this response explicit.

\subsection{Hermitianization and filter parity}

Define the Hermitian dilation
\begin{equation}
\label{eq:E18}
\mathcal A_k
=
\begin{pmatrix}
0&A_k\\
A_k^\dagger&0
\end{pmatrix}.
\end{equation}

Introduce a block-index qubit $b$, and define the controlled transition isometry
\begin{equation}
\label{eq:E19}
\overline{\square}_k
=
|0\rangle\langle0|_b\otimes\square_{P^k}
+
|1\rangle\langle1|_b\otimes\square_{(P^\star)^k}.
\end{equation}

Let
\begin{equation*}
U_k=X_b\otimes S.
\end{equation*}

Then
\begin{equation*}
\begin{aligned}
\overline{\square}_k^\dagger U_k
\overline{\square}_k
&=
\begin{pmatrix}
0&
\square_{P^k}^\dagger S\square_{(P^\star)^k}\\
\square_{(P^\star)^k}^\dagger S\square_{P^k}&0
\end{pmatrix}\\
&=
\begin{pmatrix}
0&A_k\\
A_k^\dagger&0
\end{pmatrix}.
\end{aligned}
\end{equation*}

Thus
\begin{equation}
\label{eq:E22}
\overline{\square}_k^\dagger U_k
\overline{\square}_k
=
\mathcal A_k.
\end{equation}

Let
\begin{equation*}
A_k=U\Sigma V^\dagger
\end{equation*}
be a singular-value decomposition, with complete singular-vector bases $U,V$ and nonnegative diagonal $\Sigma$.

The Hermitian dilation has paired eigenvalues $\pm\sigma_r$, where $\sigma_r$ are the singular values of $A_k$.

Let $\upsilon_k$ be the real fixed-parity filter, satisfying
\begin{equation}
\label{eq:E24}
\upsilon_k(-x)
=
(-1)^{\chi_k}\upsilon_k(x),
\qquad
\chi_k\in\{0,1\}.
\end{equation}

For even parity, the transformed dilation is block diagonal:
\begin{equation}
\label{eq:E25}
\upsilon_k(\mathcal A_k)
=
\begin{pmatrix}
U\upsilon_k(\Sigma)U^\dagger&0\\
0&V\upsilon_k(\Sigma)V^\dagger
\end{pmatrix},
\qquad \chi_k=0.
\end{equation}

For odd parity, it is block off-diagonal:
\begin{equation}
\label{eq:E26}
\upsilon_k(\mathcal A_k)
=
\begin{pmatrix}
0&U\upsilon_k(\Sigma)V^\dagger\\
V\upsilon_k(\Sigma)U^\dagger&0
\end{pmatrix},
\qquad \chi_k=1.
\end{equation}

In both cases, the positive response is
\begin{equation}
\label{eq:E27}
\upsilon_k(\mathcal A_k)^\dagger
\upsilon_k(\mathcal A_k)
=
\begin{pmatrix}
U\upsilon_k(\Sigma)^2U^\dagger&0\\
0&V\upsilon_k(\Sigma)^2V^\dagger
\end{pmatrix}.
\end{equation}

At singular value one, the corresponding left and right singular vectors are both $|\pi\rangle$. Thus the target-support projector on the Hermitianization space is
\begin{equation}
\label{eq:E28}
\Lambda_\pi^{(b)}
=
I_b\otimes\Pi_\pi.
\end{equation}

\begin{figure}[t]
\centering
\includegraphics[width=\columnwidth]{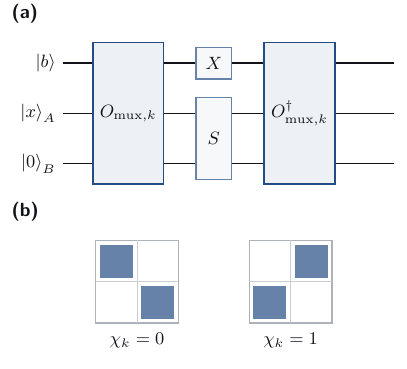}
\caption{Curved-discriminant encoding and filter parity. (a) The block-index-controlled oracle selects the forward or time-reversed powered transition; $X$ flips the block-index qubit, and $S$ swaps registers $A$ and $B$. Uncomputation and projection onto the clean transition register yield the Hermitian dilation $\mathcal A_k$. (b) The shaded blocks show even (diagonal) parity for $\chi_k=0$ and odd (off-diagonal) parity for $\chi_k=1$. Both give the same stationary target support after squaring the response.}
\label{fig:hermitianization}
\end{figure}

The response equals one on this subspace and is at most $\eta_f^2$ on its orthogonal complement. Figure~\ref{fig:hermitianization} illustrates the encoding and filter parity. 

\subsection{Complete projected-filter selective phase}
\label{subsec:projected-filter}

To use the squared response in an actual selective-phase circuit, we now specify the input and output success embeddings of a fixed-parity QSVT realization.

Let $O_{\mathrm{mux},k}$ be a unitary extension of $\overline{\square}_k$, obtained by coherently multiplexing the forward and reverse powered-transition preparations according to the block-index qubit. Define
\begin{equation*}
\mathscr U_k
=
O_{\mathrm{mux},k}^\dagger
(X_b\otimes S)
O_{\mathrm{mux},k}.
\end{equation*}

Let $A$ denote the first state register and $B$ the transition register. The isometry
\begin{equation*}
J_0:
\mathcal H_b\otimes\mathcal H_A
\longrightarrow
\mathcal H_b\otimes\mathcal H_A\otimes\mathcal H_B,
\qquad
J_0|\psi\rangle
=
|\psi\rangle|0^n\rangle_B
\end{equation*}
gives the Hermitian block encoding
\begin{equation}
\label{eq:E31}
J_0^\dagger\mathscr U_kJ_0
=
\mathcal A_k.
\end{equation}

Introduce the QSVT signal qubit $s$ and the common clean embedding
\begin{equation*}
J
=
|0\rangle_s\otimes J_0.
\end{equation*}

For the Hermitian block encoding in Eq.~\eqref{eq:E31}, let $C_k^{\mathrm{std}}$ be the standard fixed-parity QSVT realization \cite{Gilyen2019,Sunderhauf2023}, using this same clean embedding on input and output. Its accepted operator is
\begin{equation}
\label{eq:E33}
F_k^{(b)}
=
J^\dagger
C_k^{\mathrm{std}}
J
=
\upsilon_k(\mathcal A_k).
\end{equation}

More explicitly, the real-polynomial construction of Ref.~\cite{Gilyen2019} uses
\begin{equation*}
C_k^{\mathrm{std}}
=(H_s\otimes I)
\bigl[|0\rangle\langle0|_s\otimes U_\Phi
+|1\rangle\langle1|_s\otimes U_{-\Phi}\bigr]
(H_s\otimes I),
\end{equation*}
where the projected real part of the QSVT sequence $U_\Phi$ is $\upsilon_k$. The two branches have the same $d_k$ encoding queries and opposite projector phases. Their coherent multiplexing changes those phases and does not double the encoding-query count. The outer Hadamards make the signal input and successful output $s=0$, as required by $J$.

The complete success projector is
\begin{equation*}
\Pi_{\mathrm{succ}}
=
JJ^\dagger
=
|0\rangle\langle0|_s
\otimes I_{bA}
\otimes |0^n\rangle\langle0^n|_B.
\end{equation*}

By the accepted-operator convention of Appendix~\ref{app:phase-response}, the response on the input system is
\begin{equation}
\label{eq:E35}
\begin{aligned}
G_k
&=
(F_k^{(b)})^\dagger F_k^{(b)}
\\
&=
J^\dagger
(C_k^{\mathrm{std}})^\dagger
\Pi_{\mathrm{succ}}
C_k^{\mathrm{std}}
J
\\
&=
\upsilon_k(\mathcal A_k)^\dagger
\upsilon_k(\mathcal A_k).
\end{aligned}
\end{equation}

Equations~\eqref{eq:E27} and \eqref{eq:E28} show that $G_k$ commutes with $\Lambda_\pi^{(b)}$ and has
\begin{equation}
\label{eq:E36}
q_{\mathrm{tar}}=1,
\qquad
q_{\mathrm{bulk}}\leq\eta_f^2.
\end{equation}

Using the same standard circuit and success projector throughout, the complete clean-register phase circuit is
\begin{equation}
\label{eq:E38}
\widetilde S_{\phi,k}^{\mathrm{curv}}
=
(C_k^{\mathrm{std}})^\dagger
\left[
I+(e^{i\phi}-1)\Pi_{\mathrm{succ}}
\right]
C_k^{\mathrm{std}}.
\end{equation}

Finally initialize the block-index qubit in $b=0$ and define the system-to-work embedding
\begin{equation}
J_\pi|\psi\rangle
=
|0\rangle_s|0\rangle_b
|\psi\rangle_A|0^n\rangle_B.
\end{equation}

Because $\Lambda_\pi^{(b)}=I_b\otimes\Pi_\pi$, Proposition 1 and Eq.~\eqref{eq:E36} give
\begin{equation}
\label{eq:E39}
\left\|
\widetilde S_{\phi,k}^{\mathrm{curv}}J_\pi
-
J_\pi S_\phi(\Pi_\pi)
\right\|
\leq
|e^{i\phi}-1|\eta_f
\leq2\eta_f.
\end{equation}

The implementation cost is derived in Appendix~\ref{subsec:curved-phase-cost}.

\subsection{Curved phase cost and powered-scale selection}
\label{subsec:curved-phase-cost}

For the real fixed-parity filter of degree $d_k$, the construction above uses $d_k$ queries to the Hermitian block encoding $\mathscr U_k$ per filter, hence $2d_k$ per selective phase. Each encoding query contains $O_{\mathrm{mux},k}$ and its inverse. Implementing each multiplexor as a controlled forward oracle and a controlled reverse oracle gives $8d_k$ individual powered row-oracle calls per phase. The complete success phase and signal rotations are counted separately.

Each encoded query is implemented using the forward and reverse powered-transition oracles and their inverses. The primitive-oracle cost is therefore
\begin{equation*}
O\!\left(
d_k[
c_{P^k}+c_{(P^\star)^k}
]
\right).
\end{equation*}

Substituting Eq.~\eqref{eq:E17} gives
\begin{equation}
\label{eq:E41}
O\!\left(
\frac{
c_{P^k}+c_{(P^\star)^k}
}{
\sqrt{\gamma_k(P)}
}
\log\frac1{\eta_\phi}
\right).
\end{equation}

The phase-error relation in Eq.~\eqref{eq:E39} allows $\eta_f=O(\eta_\phi)$, with coefficient, rotation, and oracle errors allocated separately.

The explicit Hermitianized realization uses one block-index qubit and one QSVT signal qubit in addition to the transition-oracle registers. As in Appendix~\ref{subsec:projected-filter}, the compiled-phase cost includes the complete success projection.

For a powered-scale characterization, define the pseudo-spectral gap \cite{Paulin2015,Claudon2025}
\begin{equation}
\label{eq:E42}
\gamma_\infty(P)
=
\max_{k\geq1}\frac{\gamma_k(P)}{k}.
\end{equation}

For an ergodic finite chain, the maximum is attained at a finite integer because $\gamma_k(P)\leq1$ and $\gamma_k(P)/k\to0$ as $k\to\infty$, while sufficiently large powers of a finite ergodic kernel have strictly positive entries and therefore a positive nonstationary singular gap.

Let $k_{\mathrm{rev}}$ attain the maximum. Then
\begin{equation*}
\gamma_{k_{\mathrm{rev}}}(P)
=
k_{\mathrm{rev}}\gamma_\infty(P).
\end{equation*}

If both powered-oracle costs satisfy
\begin{equation*}
c_{P^{k_{\mathrm{rev}}}},
\,
c_{(P^\star)^{k_{\mathrm{rev}}}}
=
O(k_{\mathrm{rev}}c_P),
\end{equation*}

Eq.~\eqref{eq:E41} becomes
\begin{equation}
\label{eq:E45}
O\!\left(
c_P
\sqrt{
\frac{k_{\mathrm{rev}}}{\gamma_\infty(P)}
}
\log\frac1{\eta_\phi}
\right).
\end{equation}

%% file: appendix_F.tex
\section{Structured oracles and gate complexity}
\label{app:resources}
We derive the gate bound of Sec.~\ref{subsec:implementation-cost} and its supporting oracle, initialization, and precision costs.

\subsection{Structured Gibbs and kinetic transition oracles}

For a classical energy function $E:\Omega\to\mathbb R$, define the Gibbs distribution
\begin{equation}
\label{eq:F1}
\pi_\beta(x)
=
\frac{e^{-\beta E(x)}}{Z(\beta)},
\qquad
Z(\beta)=\sum_y e^{-\beta E(y)}.
\end{equation}

For any two configurations,
\begin{equation*}
\frac{\pi_\beta(y)}{\pi_\beta(x)}
=
e^{-\beta[E(y)-E(x)]}.
\end{equation*}

Thus a Metropolis--Hastings transition can be expressed using local energy differences and proposal probabilities. For a proposal kernel $q(x,y)$, the Metropolis--Hastings acceptance probability \cite{Hastings1970} is
\begin{equation}
\label{eq:F3}
\alpha_\beta(x,y)
=
\min\left\{
1,
e^{-\beta[E(y)-E(x)]}
\frac{q(y,x)}{q(x,y)}
\right\}.
\end{equation}

The resulting transition probability for an accepted proposal is
\begin{equation*}
Q_\beta(x,y)
=
q(x,y)\alpha_\beta(x,y)
\qquad (y\neq x),
\end{equation*}
with the rejected probability assigned to the diagonal.

A sufficient clean-access condition is that, given $x$, the distinct proposal destinations can be enumerated reversibly with bounded degree, their probabilities and energy differences can be evaluated, and the destination index can be recovered from $(x,y)$. Compute the off-diagonal probabilities $q(x,y)\alpha_\beta(x,y)$ and their sum. The remaining mass is the single diagonal probability
\begin{equation*}
Q_\beta(x,x)=1-\sum_{y\ne x}q(x,y)\alpha_\beta(x,y).
\end{equation*}

Conditional rotations then prepare the distribution on the distinct destinations together with $x$, including amplitude $\sqrt{Q_\beta(x,x)}$ on the diagonal. Map the destination index to $y$, erase the index using $(x,y)$, and uncompute the probability arithmetic. This construction combines all rejection mass before preparation and leaves no rejected-proposal record. Coherently testing acceptance while retaining the rejected proposal would not give the required oracle. Efficient proposal sampling alone is therefore insufficient; an explicit row-preparation procedure is needed. Related quantum Metropolis--Hastings circuits are discussed in Refs.~\cite{Lemieux2020,ClaudonRodenas2026}.

A structured nonreversible kernel can be obtained by composing reversible components. Suppose
\begin{equation}
\label{eq:F5}
P=QR,
\end{equation}
where $Q$ and $R$ each preserve the stationary distribution and satisfy detailed balance with respect to it. Then
\begin{equation}
\label{eq:F6}
P^\star
=
(QR)^\star
=
R^\star Q^\star
=
RQ.
\end{equation}

Kinetic samplers use velocity updates such as partial refreshment \cite{Gouraud2025}. A deterministic velocity reversal is a reversible involution when the stationary distribution is invariant under that reversal. In the eight-state example of Appendix~\ref{app:verification}, $R$ is instead a stochastic reversible velocity flip.

\subsection{Initialization and powered-transition access}

A transport procedure requires an initial coherent state \cite{AharonovTaShma2003,Ozols2012}. Let $G_{\mathrm{init}}$ be its gate-preparation cost.

For a Gibbs schedule beginning at $\beta_0=0$, the stationary distribution is uniform,
\begin{equation*}
\pi_0(x)=\frac1N,
\end{equation*}
so the initial coherent sample of the stationary distribution is
\begin{equation*}
|\pi_0\rangle
=
\frac1{\sqrt N}\sum_x|x\rangle.
\end{equation*}

For the curved route, the initial selected target is $|\pi_0\rangle$. For the flat route, the selected target is $|\mu_0\rangle$, and the initialization mismatch is measured by
\begin{equation*}
b_0
=
D_{\mathrm{tr}}(\Pi_{\pi_0},\Pi_{\mu_0}).
\end{equation*}

The costs $c_{P^m}$ and $c_{(P^\star)^m}$ use the clean endpoint-oracle definition of Eq.~\eqref{eq:A24}; the conditional linear-cost specialization is given in Eq.~\eqref{eq:A25}. The identity $P=QR$ specifies probabilities but does not by itself turn two clean component oracles into a clean endpoint oracle for their composition. In the finite example we explicitly construct the requested matrices and prepare their rows from tables. No cost $c_{P^m}=mc_P$ is assumed there.

When the initializer is known, its selective phase can be compiled directly. For the eight-state uniform state, preparation uses three Hadamard gates and its phase is $H^{\otimes3}S_\phi(|000\rangle\langle000|)H^{\otimes3}$. The same implementation is used in every comparison, without a spectral filter at the initial endpoint.

\subsection{Accuracy allocation and gate composition}

Let $n=\lceil\log_2N\rceil$, and let $g_i>0$ be the filter gap at schedule endpoint $i$:
\begin{equation}
\label{eq:F10}
g_i=
\begin{cases}
\gamma_+(Q_{i,j_i}),&\text{flat route}\\
\gamma_{k_i}(P_i),&\text{curved route}
\end{cases}
\end{equation}

As in Sec.~\ref{subsec:implementation-cost}, the gap scaling below assumes constant-factor-accurate certified bounds; otherwise the cost is governed by the supplied bounds, using $1-a_i$ for the flat filter.

Let $s_i$ be the arithmetic precision used by the controlled powered-transition oracles. Denote by $G_i(s_i)$ a gate upper bound for the required controlled clean powered-oracle operations and inverses at endpoint $i$, including their scratch-register uncomputation.

Let $R_{\mathrm{rot}}(t)$ bound the elementary-gate cost of approximating one required filter rotation to accuracy $t$ \cite{RossSelinger2016}.

The filter degree satisfies
\begin{equation}
\label{eq:F11}
d_i
=
O\!\left(
\frac1{\sqrt{g_i}}
\log\frac1{\eta_{\mathrm{ph}}}
\right),
\end{equation}

Suppose the initial mismatch and final stationary-state bias have been assigned their portions of the total error budget. Let $\epsilon_{\mathrm{alg}}>0$ denote the remaining algorithmic allowance.

Choose
\begin{equation}
\label{eq:F12}
\epsilon_i^{\mathrm{FP}}
=
\frac{\epsilon_{\mathrm{alg}}}{2\ell},
\qquad
L_{\max}=\max_iL_i.
\end{equation}

Set
\begin{equation}
\label{eq:F13}
\eta_{\mathrm{ph}}
=
\frac{
\epsilon_{\mathrm{alg}}
}{
4\ell\max\{1,L_{\max}-1\}
}.
\end{equation}

The fixed-point amplitude-amplification errors sum to
\begin{equation*}
\sum_i\epsilon_i^{\mathrm{FP}}
=
\frac{\epsilon_{\mathrm{alg}}}{2}.
\end{equation*}

Also,
\begin{equation*}
\begin{aligned}
\sum_i2(L_i-1)\eta_{\mathrm{ph}}
&\leq
2\ell\max\{1,L_{\max}-1\}\eta_{\mathrm{ph}}\\
&=
\frac{\epsilon_{\mathrm{alg}}}{2}.
\end{aligned}
\end{equation*}

Thus
\begin{equation}
\label{eq:F16}
\sum_i
\left[
2(L_i-1)\eta_{\mathrm{ph}}
+
\epsilon_i^{\mathrm{FP}}
\right]
\leq
\epsilon_{\mathrm{alg}}.
\end{equation}

The phase error is allocated among filter approximation, coefficient approximation, rotation synthesis, and oracle approximation:
\begin{equation}
\label{eq:F17}
\eta_i^{\mathrm{ph}}
\leq
\eta_{i,\mathrm{filt}}
+
\eta_{i,\mathrm{coeff}}
+
\eta_{i,\mathrm{rot}}
+
\eta_{i,\mathrm{oracle}}.
\end{equation}

A filter circuit of degree $d_i$ contains $O(d_i+1)$ oracle applications and elementary filter rotations. If the rotation approximation error is at most $t_i$ per rotation, telescoping gives an accumulated contribution $O((d_i+1)t_i)$. Therefore, a sufficient choice is
\begin{equation}
\label{eq:F18}
t_i
=
O\!\left(
\frac{\eta_{\mathrm{ph}}}{d_i+1}
\right).
\end{equation}

Likewise, if one implemented powered-oracle call differs from its ideal unitary by operator-norm error $\nu_i(s_i)$, the accumulated oracle contribution is bounded by $O((d_i+1)\nu_i(s_i))$. A sufficient precision condition is
\begin{equation}
\label{eq:F19}
(d_i+1)\nu_i(s_i)
=
O(\eta_{\mathrm{ph}}).
\end{equation}

The polynomial-completion error is controlled at the level of the filtering unitary. A uniform bound $\|\widehat C-C\|\leq\tau$ contributes at most $2\tau$ to $C^\dagger P_\phi C$. Since polynomial-coefficient errors alone need not control the completion error, one may use a certified GQSP or QSVT angle vector and bound the effect of angle rounding by telescoping. Appendix~\ref{app:verification} gives explicit sufficient tolerances for the finite construction.

The flat filter uses the quadratic filter-accuracy allocation of Eq.~\eqref{eq:D53}. The curved filter uses the linear response allocation of Eq.~\eqref{eq:E39}.

The gate cost of a compiled selective phase at endpoint $i$ is bounded conservatively by
\begin{equation}
\label{eq:F20}
\mathcal G_i
=
O\!\left(
(d_i+1)
[
G_i(s_i)+O(n)+R_{\mathrm{rot}}(t_i)
]
\right).
\end{equation}

This includes controlled powered-oracle operations, structural walk/reflection gates, and filter rotations. A constant number of filtering circuits and their inverses is absorbed into the bound.

Each transport stage uses $O(L_i)$ selective phases at its two endpoints. Including initialization, the total gate cost is
\begin{equation}
\label{eq:F21}
G_{\mathrm{total}}
=
O\!\left(
G_{\mathrm{init}}
+
\sum_{i=0}^{\ell-1}
L_i[
\mathcal G_i+\mathcal G_{i+1}
]
\right).
\end{equation}

Classical polynomial-coefficient and rotation-angle computation is accounted for separately from the quantum gate sequence.

\subsection{Space and explicit-table implementation}

Efficient quantum-walk circuits and block encodings are known for certain sparse and structured matrices \cite{ChiangNagajWocjan2010,LokeWang2017,Camps2024}, but do not directly provide the clean powered-transition oracles required here.

The two state registers of the Szegedy walk require
\begin{equation*}
2n=2\lceil\log_2N\rceil
\end{equation*}
qubits.

A complete logical-space count can be written
\begin{equation}
\label{eq:F23}
Q_{\mathrm{total}}
=
2n+
Q_{\mathrm{struct}}
+
Q_{\mathrm{oracle}}(s)
+
Q_{\mathrm{arith}}(s).
\end{equation}

The structural term includes the signal-processing and dilation registers. The flat Laurent realization uses one signal qubit for generalized quantum signal processing. The explicit curved Hermitianization uses one block-index qubit and the signal-processing registers required by its projected polynomial transformation. The oracle and arithmetic terms include the remaining workspace, including the success-projector implementation.

For a fully specified finite-matrix model, suppose certified $s$-bit approximations to every entry of each requested powered kernel are supplied in a table.

For one row, conditional probabilities can be represented by a binary tree over the padded $2^n$ destination labels. Reversible multiplexed rotations prepare the corresponding amplitude state using the conditional-probability construction of Ref.~\cite{GroverRudolph2002}. An explicit circuit applies one conditional $R_y$ rotation per binary-tree node and input row, with zero angles for branches of zero mass. Thus at most $N(2^n-1)$ such rotations suffice, each controlled on the row and previously prepared destination bits. All angles are computed classically from normalized rows. This table implementation requires no quantum arithmetic scratch register; controlling the whole oracle adds one control to each rotation. For $N=8$, the count is $56$ rotations with at most five controls, or six for a controlled oracle. These are multiple-controlled rotation counts, before decomposition into an elementary gate set.

Multiplexing over $N$ possible input rows gives the conservative gate bound
\begin{equation}
\label{eq:F24}
O\!\left(
N^2\operatorname{poly}(n,s)
\right).
\end{equation}

The classical table storage is
\begin{equation*}
O(N^2s)
\end{equation*}
bits per supplied powered kernel.

When a dense transition matrix is given explicitly, ordinary repeated squaring computes $P^m$ in
\begin{equation*}
O(N^3\log m)
\end{equation*}
arithmetic operations for $m\geq2$, in addition to precision and verified-rounding work.

Substituting Eq.~\eqref{eq:F24} into the gate-composition result gives a finite-matrix quantum-gate upper bound of the form
\begin{equation}
\label{eq:F27}
\widetilde O\!\left(
N^2\operatorname{poly}(n,s)
\sum_{i=0}^{\ell-1}
L_i[
g_i^{-1/2}+g_{i+1}^{-1/2}
]
\right),
\end{equation}
with initialization and classical preprocessing accounted for separately.

%% file: appendix_G.tex
\section{Spectral and overlap certificates}
\label{app:certificates}
We prove the overlap and spectral certificates of Sec.~\ref{subsec:parameter-certificates}.

\subsection{Gibbs-state overlap certificates}

Consider two Gibbs distributions at inverse temperatures $\beta$ and $\beta'$, with partition function
\begin{equation*}
Z(\beta)=\sum_xe^{-\beta E(x)}.
\end{equation*}

Their coherent samples satisfy
\begin{equation*}
\begin{aligned}
\langle\pi_\beta|\pi_{\beta'}\rangle
&=
\sum_x
\sqrt{
\frac{e^{-\beta E(x)}}{Z(\beta)}
\frac{e^{-\beta'E(x)}}{Z(\beta')}
}\\
&=
\frac{
Z((\beta+\beta')/2)
}{
\sqrt{Z(\beta)Z(\beta')}
}.
\end{aligned}
\end{equation*}

The squared overlap can thus be written in terms of the partition function \cite{Arunachalam2022}:
\begin{equation}
\label{eq:G3}
p(\beta,\beta')
=
\frac{
Z((\beta+\beta')/2)^2
}{
Z(\beta)Z(\beta')
}.
\end{equation}

Suppose
\begin{equation*}
E_{\max}-E_{\min}\leq B.
\end{equation*}

Set $\delta=\beta'-\beta$, and write expectations under $\pi_\beta$. Equation~\eqref{eq:G3} becomes
\begin{equation*}
p(\beta,\beta')
=
\frac{
\left(
\mathbb E_{\pi_\beta}
e^{-\delta E/2}
\right)^2
}{
\mathbb E_{\pi_\beta}e^{-\delta E}
}.
\end{equation*}

Set $T=e^{-\delta E/2}$ and choose $0<a\leq T\leq b$. Since
$(T-a)(T-b)\leq0$, writing $m=\mathbb E T$ gives
\begin{equation*}
\mathbb E T^2\leq(a+b)m-ab.
\end{equation*}

Moreover, $[(a+b)m-2ab]^2\geq0$ implies
\begin{equation*}
\frac{m^2}{(a+b)m-ab}\geq\frac{4ab}{(a+b)^2}.
\end{equation*}

Thus $p(\beta,\beta')\geq4ab/(a+b)^2$. Taking the extremal values of $T$, for which $b/a\leq e^{B|\delta|/2}$, yields
\begin{equation}
\label{eq:G10}
p(\beta,\beta')
\geq
\operatorname{sech}^{2}\!\left(\frac{B|\beta'-\beta|}{4}\right).
\end{equation}

The lower bound has quadratic small-step loss, $1-B^2\delta^2/16+O(\delta^4)$, and does not require the partition function.

For $B>0$, the sufficient temperature-step condition
\begin{equation}
\label{eq:G11}
|\beta_{i+1}-\beta_i|
\leq
\frac{4}{B}\operatorname{arcosh}\!\left(\frac{1}{\sqrt{p_{\min}}}\right)
\end{equation}
ensures $p_i\geq p_{\min}$. If $B=0$, the energy is constant and all Gibbs distributions coincide, giving $p_i=1$.

\subsection{Proof of Lemma 1: quantitative flat-route overlaps}

Under the hypotheses of Lemma 1, use the definitions of $\Delta_i$, $\gamma_i$, and $b_\star$ from Sec.~\ref{sec:resources}. The certified bias bounds imply $b_i\leq b_\star$. In particular, Eq.~\eqref{eq:D70} guarantees this whenever $\Delta_i/\gamma_i\leq b_\star^2$; the sharper bound in Eq.~\eqref{eq:D64} may be used instead.

The triangle inequality for the Fubini--Study distance then gives
\begin{equation*}
\begin{aligned}
d_{\mathrm{FS}}(\mu_i,\mu_{i+1})
&\leq\arccos\sqrt{p_i}+\arcsin b_i+\arcsin b_{i+1}\\
&\leq\arccos\sqrt{p_{\min}}+2\arcsin b_\star\\
&=\arccos\sqrt{q_{\min}}.
\end{aligned}
\end{equation*}

Taking the squared cosine yields
\begin{equation}
\label{eq:G22}
q_i\geq q_{\min}>0
\end{equation}
for every transport stage. Primitivity already implies strictly positive Perron overlaps; this estimate supplies a quantitative lower bound for the amplification cost. This proves Lemma 1. $\square$

The certified choice $\underline q_i=q_{\min}$ gives
\begin{equation}
\label{eq:G23}
L_i=O\!\left(q_{\min}^{-1/2}\log\frac1{\epsilon_i^{\mathrm{FP}}}\right).
\end{equation}

\subsection{Stage-dependent overlap certificates}

The uniform bias threshold is not necessary when adjacent stationary overlaps and spectral deficits vary along the schedule. Write $h_i=\langle\pi_i|D_i|\pi_i\rangle$. Suppose verified enclosures give
\begin{equation*}
\omega_i\in[\underline\omega_i,\overline\omega_i],\qquad
\lambda_2(D_i)\leq a_i,\qquad h_i\geq\underline h_i,
\end{equation*}
with $\underline\omega_i-a_i>0$. The sharp numerator is controlled by $\omega_i-h_i\leq\overline\omega_i-\underline h_i$, while its denominator is at least $\underline\omega_i-a_i$. If $\overline\Delta_i$ and $\underline\gamma_i>0$ also bound $\Delta_i$ and $\gamma_i$, set
\begin{equation}
\label{eq:certified-bias-min}
\overline b_i^2
=\min\!\left\{1,
\frac{\overline\omega_i-\underline h_i}{\underline\omega_i-a_i},
\frac{\overline\Delta_i}{\underline\gamma_i}\right\}.
\end{equation}

Either spectral ratio may be omitted when its inputs are unavailable. Equations~\eqref{eq:D64} and \eqref{eq:D70} give $b_i\leq\overline b_i$. The precision of the spectral enclosures must be sufficient to resolve $\omega_i-h_i$ at the scale required by the bias budget.

For a certified stationary squared-overlap bound $0<\underline p_i\leq p_i$, Eq.~\eqref{eq:B35} gives the certificate
\begin{equation}
\label{eq:stage-overlap-certificate}
\begin{aligned}
\underline q_i&=\cos^2\!\left(\min\{\pi/2,\overline\theta_i\}\right),\\
\overline\theta_i&=\arccos\sqrt{\underline p_i}+\arcsin\overline b_i
+\arcsin\overline b_{i+1}.
\end{aligned}
\end{equation}

In particular, a desired lower bound $0<q_{\min,i}\leq\underline p_i$ is certified if
\begin{equation}
\label{eq:stage-bias-budget}
\begin{aligned}
\arcsin\overline b_i+\arcsin\overline b_{i+1}
\leq{}&\arccos\sqrt{q_{\min,i}}\\
&-\arccos\sqrt{\underline p_i}.
\end{aligned}
\end{equation}

This allows different bias allowances at neighboring endpoints. Whenever the bound in Eq.~\eqref{eq:stage-overlap-certificate} is positive, it can be used directly in Corollary 2. The leading-eigenvalue filtering hypotheses and the initial and final bias budget must still be checked separately.

\subsection{Direct Perron-state overlap certificate}

Let $\Pi_{i+1}=|\mu_{i+1}\rangle
\langle\mu_{i+1}|$ and
$\Pi_{i+1}^{\perp}=I-\Pi_{i+1}$.
The Perron eigenvalue equation gives
\begin{equation*}
(D_{i+1}-\omega_i I)|\mu_i\rangle
=
(D_{i+1}-D_i)|\mu_i\rangle.
\end{equation*}

Since $\Pi_{i+1}^{\perp}$ commutes with $D_{i+1}$,
projecting onto the orthogonal complement of
$|\mu_{i+1}\rangle$ gives
\begin{equation*}
\begin{aligned}
&\|(D_{i+1}-\omega_i I)
\Pi_{i+1}^{\perp}|\mu_i\rangle\|\\
&\qquad\leq\|D_{i+1}-D_i\|.
\end{aligned}
\end{equation*}

If $w_i>a_{i+1}$, the spectral theorem implies
\begin{equation*}
(w_i-a_{i+1})
\|\Pi_{i+1}^{\perp}|\mu_i\rangle\|
\leq\overline e_i.
\end{equation*}

Using
$\|\Pi_{i+1}^{\perp}|\mu_i\rangle\|^2=1-q_i$
gives
\begin{equation*}
1-q_i\leq
\frac{\overline e_i^2}{(w_i-a_{i+1})^2}.
\end{equation*}

Interchanging $i$ and $i+1$ gives the corresponding
bound in the opposite direction. Taking the larger
positive spectral separation proves
Eq.~\eqref{eq:direct-overlap-bound}. $\square$

For explicitly known discriminants, a sufficient
bound on their variation is
\begin{equation*}
\|D_{i+1}-D_i\|
\leq
\max_x\sum_y
|D_{i+1}(x,y)-D_i(x,y)|.
\end{equation*}

For larger structured systems, such bounds must be
supplied analytically or through a separate
certification procedure. They are not assumed to
follow efficiently from coherent transition access
alone.

\subsection{Spectral certification for the two routes}

The curved construction requires a lower bound on the powered singular gap. Let
\begin{equation*}
A_{i,k_i}
=
D\!\left(
P_i^{k_i},(P_i^\star)^{k_i}
\right).
\end{equation*}

If a certified enclosure gives
\begin{equation*}
s_2(A_{i,k_i})\leq r_i<1,
\end{equation*}
then Eq.~\eqref{eq:E14} gives
\begin{equation}
\label{eq:G31}
\gamma_{k_i}(P_i)
\geq1-r_i^2.
\end{equation}

For the flat construction, certify primitivity and spectral bounds
$\lambda_2(D_i)\leq a_i<1$ and $\omega_i\geq w_i>0$. The interval filter of Appendix~\ref{app:flat} is applicable when
\begin{equation}
\label{eq:operational-filter-premise}
-1<a_i<1,\qquad 1-w_i\leq\frac{1-a_i}{16}.
\end{equation}

Its interval parameter is $\delta_i=1-a_i\in(0,2)$. Negative $a_i$ is therefore allowed without shrinking the acceptance interval. Together with a certified lower bound on $h_i$, these enclosures yield Eq.~\eqref{eq:certified-bias-min}; stationary-overlap bounds then yield Eq.~\eqref{eq:stage-overlap-certificate}. Filtering admissibility and final stationary-state accuracy are separate checks.

For explicitly known finite matrices, directed interval evaluation of the transition rules, normalization, matrix products, and square roots controls the entries of $D_i$ and $h_i$. Verified eigenvalue bounds then supply the remaining inputs; Appendix~\ref{app:verification} instantiates this procedure for eight states. Matrix construction, normalization, and spectral certification contribute to the classical preprocessing cost. Clean powered row-state preparations, their inverses, and their controlled versions must also be supplied as in Appendix~\ref{app:resources}.

For structured larger systems, analytic estimates of contraction, conductance, spectral separation, or mixing may supply these certificates. Their validity and parameter-generation costs depend on the family of kernels.

\subsection{Cooling-schedule implications}

Annealing and sequential Monte Carlo use sequences of intermediate distributions \cite{Neal2001,WillsSchon2023}. Shifting integer-valued energies by their minimum leaves every Gibbs distribution unchanged and ensures $Z(\infty)\geq1$. We may therefore write
\begin{equation*}
H:\Omega\to\{0,\ldots,h\},
\qquad
\min_x H(x)=0,
\qquad
h\geq2.
\end{equation*}

Following earlier adaptive cooling schedules \cite{HarrowWei2020,StefankovicVempalaVigoda2009}, the balanced-schedule bound of Ref.~\cite{Arunachalam2022} (Theorem 2.2) and its overlap identity (Fact 3.1) yield, for any finite target inverse temperature,
\begin{equation*}
\ell=O\!\left(1+\sqrt{\ln|\Omega|\,\ln h}\right)
\end{equation*}
with constant adjacent squared overlaps. Indeed, writing $g=Z(\infty)$, while $Z(\beta)\geq2g$ we have $1/2\leq-(\ln Z)'\leq h$, so the derivative ratio in the schedule bound is at most $2h$. Once $Z(\beta)\leq2g$, any two colder QSAMPLEs have ground-state mass at least $1/2$ and squared overlap at least $1/4$; one final step suffices. The zero-temperature limit need not satisfy the full-support assumption.

In the curved route, such a stationary overlap lower bound is directly the selected-state overlap lower bound.

In the flat route, it combines with the bias estimates of Appendix~\ref{app:flat} and the sufficient conditions of Lemma 1 to yield quantitative Perron-state overlap bounds.

%% file: appendix_H.tex
\section{Finite-dimensional verification and benchmarks}
\label{app:verification}
We give the complete finite-dimensional calculations supporting Sec.~\ref{sec:verification} and the winning-streak benchmark.

\subsection{Eight-state nonreversible Gibbs model}

Consider
\begin{equation}
\label{eq:H1}
\Omega=\mathbb Z_4\times\{+,-\}.
\end{equation}

The configuration coordinate is $x\in\{0,1,2,3\}$, and $v\in\{+,-\}$ denotes a velocity variable.

Set
\begin{equation}
\label{eq:H2}
E=(0,0.4,1.2,0.8).
\end{equation}

At inverse temperature $\beta$,
\begin{equation}
\label{eq:H3}
\pi_\beta(x,v)
=
\frac{e^{-\beta E_x}}{2Z(\beta)},
\qquad
Z(\beta)=\sum_{x=0}^3e^{-\beta E_x}.
\end{equation}

For a fixed velocity $v$, propose a clockwise step $x\mapsto x+1\pmod4$ with probability
\begin{equation*}
a_+=\frac34,
\qquad
a_-=\frac14,
\end{equation*}
and a counterclockwise step with probability $1-a_v$.

Write $q_v(x,y)$ for this proposal distribution. The Metropolis acceptance probability is
\begin{equation}
\label{eq:H5}
\alpha_{\beta,v}(x,y)
=
\min\left\{
1,
e^{-\beta(E_y-E_x)}
\frac{q_v(y,x)}{q_v(x,y)}
\right\}.
\end{equation}

The kernel $Q_\beta$ retains the velocity label. Its accepted transitions are
\begin{equation}
\label{eq:H6}
Q_\beta((x,v),(y,v))
=
q_v(x,y)\alpha_{\beta,v}(x,y)
\qquad(y\neq x),
\end{equation}
and its diagonal receives the rejection probability,
\begin{equation}
\label{eq:H7}
Q_\beta((x,v),(x,v))
=
1-
\sum_{y\neq x}
q_v(x,y)\alpha_{\beta,v}(x,y).
\end{equation}

Define the independent velocity-flip kernel
\begin{equation}
\label{eq:H8}
R((x,v),(x,-v))=\frac25,
\end{equation}
\begin{equation}
\label{eq:H9}
R((x,v),(x,v))=\frac35.
\end{equation}

Both $Q_\beta$ and $R$ satisfy detailed balance with respect to the distribution in Eq.~\eqref{eq:H3}. Their composition defines
\begin{equation}
\label{eq:H10}
P_\beta=Q_\beta R,
\qquad
P_\beta^\star=RQ_\beta.
\end{equation}

Thus $\pi_\beta$ is stationary for both kernels, and their ordered compositions generally differ.

\subsection{Certified spectra, biases, and overlaps}

We use $\beta=0,0.75,1.25$. The staged flat paths have intermediate power $j=4$ or $1$ and the same final power-$4$ target. The direct flat path omits the intermediate stage, while the direct curved path uses $k=1$. All start from the uniform state; in each staged path take $j_0=j_1$.

We evaluate the transition rules with directed interval arithmetic at 85 decimal digits. Interval $LDL^{\mathsf T}$ inertia tests certify eigenvalue counts above rational endpoints. The internal eigenvalue enclosures have width at most $4\times10^{-36}$. A strictly positive seventh power of each required flat discriminant certifies primitivity; the $\beta=0$ case is exact, as shown below.

\begin{table*}[t]
\caption{Conservative spectral certificates. Here $\omega\geq w$, $\lambda_2\leq a$, $\underline n\leq\omega-\langle\pi|D|\pi\rangle\leq\overline n$, and $\omega-\lambda_2\geq\underline g$. Equation~\eqref{eq:flat-bias-bound} gives $\underline b\leq b\leq\overline b$. The bounds $a$ and $w$ also determine the filter parameters.}
\label{tab:gibbs-data}
\begin{ruledtabular}
\begin{tabular}{ccccccccc}
$\beta$ & $j$ & $w$ & $a$ & $\underline n$ & $\overline n$ & $\underline g$ & $\underline b$ & $\overline b$\\
\hline
$0.75$ & $1$ & $0.9913$ & $0.489$ & $2.613\times10^{-5}$ & $2.614\times10^{-5}$ & $0.5029$ & $3.614\times10^{-3}$ & $7.21\times10^{-3}$\\
$0.75$ & $4$ & $0.99994$ & $0.059$ & $0$ & $9.881\times10^{-10}$ & $0.9410$ & $0$ & $3.25\times10^{-5}$\\
$1.25$ & $4$ & $0.99991$ & $0.053$ & $0$ & $4.047\times10^{-9}$ & $0.9470$ & $0$ & $6.54\times10^{-5}$\\
\end{tabular}
\end{ruledtabular}
\end{table*}

The numerator and denominator of Eq.~\eqref{eq:flat-bias-bound} are enclosed separately to resolve the small numerator. For comparison, the numerical deficit-to-gap ratios at $j=4$ are $5.50\times10^{-5}$ and $8.99\times10^{-5}$ at the two nonzero inverse temperatures. The coarse bias bound would therefore be insufficient for the chosen endpoint allowance. All rows of Table~\ref{tab:gibbs-data} satisfy $1-w\leq(1-a)/16$. The curved calculation encloses the second eigenvalue of $A_1A_1^\dagger$, giving $s_2(A_1)<0.513$ and $\gamma_1(P_{5/4})>0.73727$.

With $B=6/5$, Eq.~\eqref{eq:gibbs-overlap-bound} gives stationary-overlap lower bounds $0.951035$, $0.977833$, and $0.871579$ for the two staged transitions and the direct transition, respectively. Combining the analytic range bounds with Table~\ref{tab:gibbs-data} in Eq.~\eqref{eq:overlap-bound} gives the downward-rounded bounds in Table~\ref{tab:certified-comparison}. The actual numerical estimates for the power-$4$ path are plotted in Fig.~\ref{fig:gibbs-schedule}; for the path with $j_1=1$ the overlaps are approximately $0.971034$ and $0.990243$, and for direct flat transport the overlap is $0.929019$.

For the tight geometry bounds in Table~\ref{tab:overlap-comparison}, interval evaluation of the stationary overlaps from Eq.~\eqref{eq:H3} is combined with the upper biases in Table~\ref{tab:gibbs-data} using Eq.~\eqref{eq:overlap-bound}. For the direct certificate, the inputs in Table~\ref{tab:direct-inputs} use only the discriminants. Interval inertia tests at $\pm\overline e_i$ enclose the eigenvalues of $D_{i+1}-D_i$ in $[-\overline e_i,\overline e_i]$. The cross-gaps use Table~\ref{tab:gibbs-data} and the exact initial eigenvalues $\omega=1$, $\lambda_2=2^{-j_0}$.

\begin{table}[ht]
\caption{Inputs to Lemma 2 for the staged paths. $\overline e_i$ is an upper bound on $\|D_{i+1}-D_i\|$, and $g_i^{\rm cross}$ is the certified spectral cross-gap used in Eq.~\eqref{eq:direct-overlap-bound}.}
\label{tab:direct-inputs}
\begin{ruledtabular}
\begin{tabular}{cccc}
$j_1$ & Step & $\overline e_i$ & $g_i^{\rm cross}$ (lower bound)\\
\hline
$1$ & First & $0.1801$ & $0.511$\\
$1$ & Second & $0.4530$ & $0.9383$\\
$4$ & First & $0.1643$ & $0.941$\\
$4$ & Second & $0.1111$ & $0.94694$\\
\end{tabular}
\end{ruledtabular}
\end{table}

\subsection{Explicit filters and final-accuracy allocation}

The common error target and its allocation are specified in Sec.~\ref{sec:verification}. Initialization and its direct selective phase are given in Appendix~\ref{app:resources}. For each transition we use the smallest odd $L_i$ satisfying Eq.~\eqref{eq:explicit-fpaa-length}, with $\epsilon_i^{\mathrm{FP}}=2.25\times10^{-4}$ for staged transitions and $4.5\times10^{-4}$ for direct transport. All $L_i=5$ except for the $j_1=1$ direct-overlap construction, which uses $L_0=L_1=7$. Each step uses $(L_i-1)/2$ phases about each endpoint. The conservative phase-error allowance is
\begin{equation}
\eta_{\mathrm{ph}}=\frac{4.5\times10^{-4}}{2\sum_i(L_i-1)},
\label{eq:H-phase-budget}
\end{equation}
whose denominator is $16$ for the $L_i=5$ staged paths, $24$ for the $L_i=7$ staged path (giving $\eta_{\rm ph}=1.875\times10^{-5}$), and $8$ for direct transport.

For each flat filter, use the rational bounds $a,w$ in Table~\ref{tab:gibbs-data}. Put $\delta=1-a$, $m=\lceil\delta^{-1/2}\rceil=2$, and use the squared Chebyshev base $u$ of Appendix~\ref{app:flat}. Choose the smallest odd $h$ for which the two exact binomial tails
\begin{equation*}
B_h(u_{
\rm bulk})\leq\eta_f,
\qquad 1-B_h(u(w))\leq\eta_f,
\end{equation*}
are certified, where $u_{\rm bulk}=1/T_m((2+\delta)/(2-\delta))^2$ and $\eta_f=\kappa=\eta_{\mathrm{ph}}^2/256$. The implemented polynomial is $(1-\kappa)B_h(u(x))$, of degree $d=2mh$. This specifies its rational coefficients exactly; no coefficient rounding is required. The staged $j=4$ filters have $h=13$, the intermediate $j=1$ filter has $h=27$ for $L_i=5$ and $h=29$ for $L_i=7$, and the direct final flat filter has $h=11$. Thus the $L_i=7$ path uses $d=(116,52)$. Equation~\eqref{eq:D52} allocates at most $\eta_{\mathrm{ph}}/4$ to the filter response error.

For the curved baseline use the even polynomial
\begin{equation}
\upsilon_d(x)=\frac{T_d(x/r)}{T_d(1/r)},\qquad r=0.513,
\label{eq:H-curved-polynomial}
\end{equation}
with the smallest positive even $d$ satisfying $1/T_d(1/r)\leq\eta_{\mathrm{ph}}/8$. This gives $d=10$. It is bounded by one on $[-1,1]$, equals one at both endpoints, and has magnitude at most $\eta_{\mathrm{ph}}/8$ throughout $[-r,r]$. The resulting phase-response error is at most $\eta_{\mathrm{ph}}/4$.

The remaining three quarters of each phase budget are allocated equally to polynomial completion, rotation synthesis, and row-oracle approximation. Sufficient explicit tolerances are
\begin{equation}
\begin{aligned}
\|\widehat C-C\|&\leq\eta_{\mathrm{ph}}/8,\\
t_{\rm rot}&\leq\frac{\eta_{\mathrm{ph}}}{4(12d+9)},\\
\nu_{\rm row}&\leq\frac{\eta_{\mathrm{ph}}}{4M_{\rm row}},
\end{aligned}
\label{eq:H-precision}
\end{equation}
where $M_{\rm row}=8d+2$ for flat phases and $8d$ for curved phases. The first condition applies to each completed filter before gate synthesis, giving at most $2\|\widehat C-C\|$ in filter--phase--uncompute. The second is a conservative tolerance for each programmable signal or projector-phase rotation, allowing three Euler rotations per signal-processing rotation; any decomposition of multiple controls must meet the same total norm tolerance. The third includes every preparation and unpreparation in the phase; assigning each of the $56$ row-preparation rotations norm error at most $\nu_{\rm row}/56$ suffices. A direct initializer phase is required to have norm error at most $\eta_{\mathrm{ph}}$. These are sufficient implementation requirements; GQSP/QSVT phase synthesis and compiled-circuit simulation are not performed.

The certified endpoint bound and these budgets give
\begin{equation*}
D_{\mathrm{tr}}(\rho_2,\Pi_{\pi_2})
\leq9.66\times10^{-4}
\end{equation*}
for every flat configuration, and at most $9\times10^{-4}$ for the curved baseline, for implementations meeting Eq.~\eqref{eq:H-precision}.

\subsection{Access model and resource counts}

The explicit-table model of Appendix~\ref{app:resources} uses $56$ multiple-controlled rotations per clean eight-row oracle, with inversion and external controls implemented on those rotations. Every powered row-oracle call has the same cost in this model. Classical powering affects the preprocessing and spectrum, not the quantum query cost.

One flat phase uses $4d$ controlled walk applications and $8d+2$ row-oracle calls; one curved phase uses $2d$ Hermitian block queries and $8d$ row-oracle calls. Staged transport uses four intermediate and two final phases for $L_i=5$, or six intermediate and three final phases for $L_i=7$. Direct transport uses two final phases. Initializer phases use no row oracles. Hence the counts in Table~\ref{tab:certified-comparison} are
\begin{equation*}
\begin{aligned}
N_O^{\mathrm{staged}}&=4(8d_1+2)+2(8d_2+2),\\
N_O^{\mathrm{staged},L=7}&=6(8d_1+2)+3(8d_2+2)=6834,\\
N_O^{\mathrm{direct\ flat}}&=2(8d+2),\\
N_O^{\mathrm{direct\ curved}}&=16d.
\end{aligned}
\end{equation*}

The flat walk counts are $1248$ for $j_1=4$, $2144$ for $j_1=1$ with $L_i=5$, $3408$ for $j_1=1$ with $L_i=7$, and $352$ for direct flat transport; the curved construction uses $40$ block queries. Multiplying $N_O$ by $56$ bounds the number of multiple-controlled row rotations, excluding signal rotations, structural gates and initializer phases. The explicit realizations use seven flat or eight curved register qubits before additional synthesis workspace. These are not optimized elementary-gate counts. Classical preprocessing, interval tests and polynomial completion are accounted for separately; the Gibbs table also permits direct amplitude-state preparation, so the comparison establishes no advantage over all available methods.

\subsection{Exact reference-case and operator checks}
\label{subsec:exact-reference-case}

At $\beta=0$, the biased proposals in either velocity sector yield the same lazy position walk: the accepted clockwise and counterclockwise probabilities are $1/4$ each, with holding probability $1/2$. Its circulant spectrum is $1,1/2,1/2,0$. The independent velocity-flip matrix has spectrum $1,1/5$. Thus $P_0$ is symmetric with the product spectrum, and $D_{\mathrm{flat}}(P_0^4)=P_0^4$ has second eigenvalue $1/16$. Consequently,
\begin{equation*}
\omega_4(0)=1,\quad \gamma_+(Q_{0,4})=15/16,\quad \Delta_4(0)=b_4(0)=0.
\end{equation*}

For the one-step kernel $P_{1.25}$, numerical checks reproduce the discriminant-encoding and conjugate-pair projector compression identities to within $10^{-13}$. Explicit Hermitianization also reproduces the parity patterns of Eqs.~\eqref{eq:E25} and \eqref{eq:E26} and the squared-response support of Eq.~\eqref{eq:E27}.

\subsection{Uniform temperature certificate}

For $0\leq\beta\leq0.75$, the maximum neighboring energy difference is $0.8$. In each row of $Q_\beta$, one accepted position-transition probability is $1/4$ and the other is $(1/4)e^{-\beta\Delta E}$ with $|\Delta E|\leq0.8$: the latter remains below the larger proposal probability $3/4$. The holding probability is $3/4-(1/4)e^{-\beta\Delta E}$. Each nonzero transition has logarithmic derivative bounded in magnitude by
\begin{equation*}
\max\left\{0.8,\frac{0.2e^{0.6}}{0.75-0.25e^{0.6}}\right\}<\frac54.
\end{equation*}
Since $R$ is independent of temperature, the bound carries over to $P_\beta$ and $D_\beta$. Their support is fixed, with the symmetric discriminant support connected and its diagonal positive, so $D_\beta$ is primitive. Entrywise, $|\partial_\beta D_\beta|\leq(5/4)D_\beta$. For any real unit vector $v$, bounding the quadratic form by $(5/4)|v|^{\mathsf T}D_\beta|v|$ gives $\|\partial_\beta D_\beta\|\leq5/4$ because $\|D_\beta\|\leq1$. Integration proves the Lipschitz bound in Sec.~\ref{sec:verification}.

Interval inertia tests at $0.991$ and $0.501$ certify the leading and second-eigenvalue bounds at each of the 26 anchors (the $\beta=0$ spectrum is exact). Weyl's inequality extends these bounds to the interval: $w=0.991-(5/4)0.015=0.97225$, $a=0.501+(5/4)0.015=0.51975$. Here $1-w=0.02775<(1-a)/16$ and the cross-gap is at least $w-a=0.4525$. Substituting into Lemma 2 proves Eq.~\eqref{eq:uniform-overlap}.

\subsection{Independent copies}

For $M$ independent copies, $P_{\beta,M}=P_\beta^{\otimes M}$ and $D_{\beta,M,j}=D_{\beta,j}^{\otimes M}$. The Perron and stationary states factorize, giving
\begin{equation*}
\omega_M=\omega^M,\quad q_{i,M}=q_{i,1}^M,\quad
b_{i,M}=\sqrt{1-(1-b_{i,1}^2)^M}.
\end{equation*}

For $M\geq2$, put $r_+=\lambda_2(D)/\omega$, $r_-=\max\{0,-\lambda_{\min}(D)\}/\omega$, and $s=\max\{r_+,r_-^2\}$. Products of the normalized eigenvalues give $\gamma_+(Q_M)=1-s$: a positive nontarget factor is at most $r_+$, and products of two negative factors are at most $r_-^2$. The filter premise becomes $\omega^M\geq15/(16-s)$ and eventually fails at fixed power if $\omega<1$. Although direct overlap can shrink with $M$, the stationary QSAMPLE also admits factorwise preparation, so the product construction alone does not show a speedup.

\subsection{Winning-streak benchmark and powered transitions}

Consider the winning-streak chain \cite{LevinPeres2017} on
\begin{equation*}
\Omega_N=\{0,\ldots,N\}.
\end{equation*}

At each step, the chain resets to $0$ with probability $1/2$. Otherwise it advances from $x$ to $\min(x+1,N)$.

Thus
\begin{equation*}
P_N(x,0)=\frac12,
\end{equation*}
and
\begin{equation*}
P_N(x,\min(x+1,N))=\frac12,
\end{equation*}
with coincident destination probabilities added where applicable.

Its stationary distribution is
\begin{equation}
\label{eq:H39}
\pi(0)=\frac12,
\end{equation}
\begin{equation}
\label{eq:H40}
\pi(x)=2^{-(x+1)}
\qquad(1\leq x<N),
\end{equation}
and
\begin{equation}
\label{eq:H41}
\pi(N)=2^{-N}.
\end{equation}

These probabilities sum to one. Stationarity follows from the reset-and-advance structure: the last reset determines the position, with the geometric tail accumulated at $N$.

The forward chain mixes in a number of steps independent of $N$ at fixed total-variation tolerance. Coupling two copies with the same reset/advance choices, the copies coalesce at their first simultaneous reset. The probability of no reset during the first $t$ steps is $2^{-t}$, giving
\begin{equation}
\label{eq:H42}
d_{\mathrm{TV}}(P_N^t(x,\cdot),\pi)
\leq2^{-t}.
\end{equation}

Hence the worst-start mixing time is $O(\log(1/\epsilon))$, independently of $N$.

For the time reversal $P_N^\star$, the chain started at $N$ remains there with probability $1/2$ or moves to $N-1$ with probability $1/2$. From an interior state $x\in\{1,\ldots,N-1\}$, it moves deterministically to $x-1$. Thus a worst-start trajectory requires order $N$ steps to reach the region carrying most stationary mass. At fixed nontrivial total-variation tolerance, its worst-start mixing time is of order $N$.

The reverse chain has a different behavior from the initial state $0$. Since $P_N(y,0)=1/2$ for every $y$ and $\pi(0)=1/2$,
\begin{equation*}
P_N^\star(0,y)
=
\frac{\pi(y)}{\pi(0)}P_N(y,0)
=\pi(y).
\end{equation*}

Therefore, $P_N^\star$ started at $0$ reaches the stationary distribution in one step.

For any kernel with stationary distribution $\pi$,
\begin{equation*}
(P^\star)^j(x,y)
=
\frac{\pi(y)}{\pi(x)}P^j(y,x).
\end{equation*}

Consequently,
\begin{equation}
\label{eq:H45}
\begin{aligned}
D_{\mathrm{flat}}((P^\star)^j)(x,y)
&=
\sqrt{(P^\star)^j(x,y)(P^\star)^j(y,x)}\\
&=
\sqrt{P^j(y,x)P^j(x,y)}\\
&=
D_{\mathrm{flat}}(P^j)(x,y).
\end{aligned}
\end{equation}

Thus the forward and reverse winning-streak chains have the same powered flat discriminant.

For $1\leq j\leq N$, the powered transitions have the following closed form; the plotted $N=16,j=20$ case is evaluated by direct matrix powering.
\begin{equation}
\label{eq:H46}
P_N^j(x,y)
=
\sum_{t=0}^{j-1}
2^{-(t+1)}\mathbf1_{\{y=t\}}
+
2^{-j}
\mathbf1_{\{y=\min(x+j,N)\}}.
\end{equation}

To derive it, distinguish paths by their last reset during the $j$ steps. A last reset followed by $t$ advances gives final state $t$ and probability $2^{-(t+1)}$ for $0\leq t<j$. A path without a reset has probability $2^{-j}$ and ends at $\min(x+j,N)$.

The two terms have disjoint support because
\begin{equation*}
\min(x+j,N)\geq j
\end{equation*}
for $j\leq N$.

Equation~\eqref{eq:H46} also gives an explicit structured powered-oracle construction. The first term is a truncated geometric distribution, and the second is a point mass. Their dyadic interval masses, meaning the probabilities of intervals in a binary partition, can be computed from geometric sums. Reversible conditional-rotation state preparation \cite{GroverRudolph2002} therefore gives a clean endpoint oracle with gate cost polynomial in $\log(N+1)$ and the required arithmetic precision for this structured family.

\begin{figure}[t]
\centering
\includegraphics[width=\columnwidth]{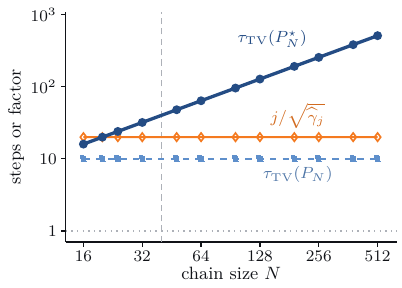}
\caption{Winning-streak mixing and query-cost proxy ($j=20$). At total-variation tolerance $10^{-3}$, worst-start mixing takes $10$ steps for $P_N$ and $N$ steps for $P_N^\star$; starting $P_N^\star$ at $0$ reaches stationarity in one step (dotted line). The factor $j/\sqrt{\widehat\gamma_j}$ is a query proxy, not a measured gate cost. The dashed vertical line marks $N=2j=40$, where the flat discriminant becomes nonprimitive and $\widehat\gamma_j$ is only a spectral-separation diagnostic.}
\label{fig:winning-mixing}
\end{figure}

\begin{figure}[t]
\centering
\includegraphics[width=\columnwidth]{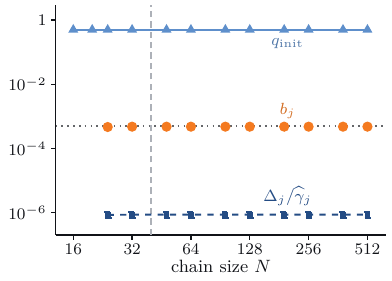}
\caption{Winning-streak spectral diagnostics ($j=20$). Shown are the stationary-to-Perron trace bias $b_j$ (markers), normalized deficit $\Delta_j/\widehat\gamma_j$, and initial squared overlap $q_{\mathrm{init}}=|\langle0|\mu_j\rangle|^2\simeq0.5$. Both $b_j$ and $\Delta_j/\widehat\gamma_j$ vanish at $N=16,20$ and are omitted from the logarithmic axis. The dotted horizontal line gives the $5\times10^{-4}$ bias allowance; the dashed vertical line marks the nonprimitivity threshold $N=2j=40$.}
\label{fig:winning-spectral}
\end{figure}

\subsection{Spectral diagnostics, primitivity, and initialization}

For the winning-streak benchmark, define
\begin{equation}
\label{eq:H48}
D_j
=
D_{\mathrm{flat}}(P_N^j)
=
D_{\mathrm{flat}}((P_N^\star)^j).
\end{equation}

Let $\omega_j$ and $\lambda_2(D_j)$ be its largest and second-largest eigenvalues. The normalized spectral-separation diagnostic is
\begin{equation}
\label{eq:H49}
\widehat\gamma_j
=
1-\frac{\lambda_2(D_j)}{\omega_j}.
\end{equation}

When $D_j$ is primitive, this equals the geometric-reversibilization gap $\gamma_+(Q_j)$.

For $N\geq2j$, the powered flat discriminant is reducible. Indeed, Eq.~\eqref{eq:H46} shows that $P_N^j(x,y)>0$ only when $y<j$ or $y=\min(x+j,N)$.

For states $2j\leq x<N$, the possible transitions are not bidirectional, so
\begin{equation}
\label{eq:H52}
D_j(x,y)=0
\qquad\text{for every }y.
\end{equation}

The endpoint $N$ is also isolated in the discriminant, with diagonal entry $D_j(N,N)=2^{-j}$.

Thus the discriminant is nonprimitive for $N\geq2j$. In that regime, $\widehat\gamma_j$ is the normalized spectral-separation diagnostic defined in Eq.~\eqref{eq:H49}.

The benchmark in Fig.~\ref{fig:winning-mixing} uses $j=20$ and evaluates the worst-start $10^{-3}$ total-variation mixing times of the forward and reverse chains. The mixing times are compared with the illustrative query-cost proxy $j/\sqrt{\widehat\gamma_j}$.

The diagnostics in Fig.~\ref{fig:winning-spectral} examine the selected Perron state $|\mu_j\rangle$. Its stationary-state trace bias is
\begin{equation}
\label{eq:H55}
b_j
=
\sqrt{1-|\langle\pi|\mu_j\rangle|^2},
\end{equation}
and the normalized deficit is
\begin{equation}
\label{eq:H56}
\frac{\Delta_j}{\widehat\gamma_j},
\qquad
\Delta_j
=
1-\langle\pi|D_j|\pi\rangle.
\end{equation}

For a one-stage initialization from $|0\rangle$, the selected-state squared overlap is
\begin{equation}
\label{eq:H57}
q_{\mathrm{init}}
=
|\langle0|\mu_j\rangle|^2.
\end{equation}

The numerical diagnostics in Figs.~\ref{fig:winning-mixing} and \ref{fig:winning-spectral} use ordinary floating-point arithmetic and give $q_{\mathrm{init}}\simeq1/2$, so the overlap-dependent initialization factor is of constant order.

The stationary distribution in Eqs.~\eqref{eq:H39}--\eqref{eq:H41} also has a direct coherent preparation construction. Its dyadic interval masses are geometric sums, enabling conditional-rotation preparation with cost polynomial in $\log(N+1)$ and the arithmetic precision.

The benchmark therefore separates several quantities that have distinct algorithmic roles: classical mixing from a specified or worst-case initialization, normalized flat spectral separation, stationary-target bias, and overlap with the chosen initial state.

%% file: references.bib
@inproceedings{ChenLovaszPak1999,
  author        = {Chen, F. and Lov\'asz, L. and Pak, I.},
  title         = {Lifting {Markov} chains to speed up mixing},
  booktitle     = {Proceedings of the 31st Annual ACM Symposium on Theory of Computing},
  publisher     = {ACM},
  address       = {New York},
  pages         = {275--281},
  year          = {1999},
  doi           = {10.1145/301250.301315}
}

@article{DiaconisHolmesNeal2000,
  author        = {Diaconis, P. and Holmes, S. and Neal, R. M.},
  title         = {Analysis of a nonreversible {Markov} chain sampler},
  journal       = {Ann. Appl. Probab.},
  volume        = {10},
  pages         = {726--752},
  year          = {2000},
  doi           = {10.1214/aoap/1019487508}
}

@article{Paulin2015,
  author        = {Paulin, D.},
  title         = {Concentration inequalities for {Markov} chains by {Marton} couplings and spectral methods},
  journal       = {Electron. J. Probab.},
  volume        = {20},
  pages         = {79},
  year          = {2015},
  doi           = {10.1214/EJP.v20-4039}
}

@article{Somma2008,
  author        = {Somma, R. D. and Boixo, S. and Barnum, H. and Knill, E.},
  title         = {Quantum simulations of classical annealing processes},
  journal       = {Phys. Rev. Lett.},
  volume        = {101},
  pages         = {130504},
  year          = {2008},
  doi           = {10.1103/PhysRevLett.101.130504}
}

@article{MagniezNayakRolandSantha2011,
  author        = {Magniez, F. and Nayak, A. and Roland, J. and Santha, M.},
  title         = {Search via quantum walk},
  journal       = {SIAM J. Comput.},
  volume        = {40},
  pages         = {142--164},
  year          = {2011},
  doi           = {10.1137/090745854}
}

@article{OrfiSels2024,
  author        = {Orfi, A. and Sels, D.},
  title         = {Bounding the speedup of the quantum-enhanced {Markov}-chain {Monte Carlo} algorithm},
  journal       = {Phys. Rev. A},
  volume        = {110},
  pages         = {052414},
  year          = {2024},
  doi           = {10.1103/PhysRevA.110.052414}
}

@misc{GroverRudolph2002,
  author        = {Grover, L. and Rudolph, T.},
  title         = {Creating superpositions that correspond to efficiently integrable probability distributions},
  year          = {2002},
  eprint        = {quant-ph/0208112},
  archivePrefix = {arXiv}
}

@article{Claudon2025,
  author        = {Claudon, B. and Piquemal, J.-P. and Monmarch\'e, P.},
  title         = {Quantum speedup for nonreversible {Markov} chains},
  journal       = {Nat. Commun.},
  volume        = {16},
  pages         = {10732},
  year          = {2025},
  doi           = {10.1038/s41467-025-65761-5},
  note          = {See also \href{https://pmc.ncbi.nlm.nih.gov/articles/instance/12663442/bin/41467_2025_65761_MOESM1_ESM.pdf}{Supplementary Information}}
}

@misc{Zhao2026,
  author        = {Zhao, N.},
  title         = {Ancilla-efficient {QSAMPLE} preparation for reversible {Markov} chains},
  year          = {2026},
  eprint        = {2605.23442v2},
  archivePrefix = {arXiv},
  primaryClass  = {quant-ph}
}

@article{MotlaghWiebe2024,
  author        = {Motlagh, D. and Wiebe, N.},
  title         = {Generalized quantum signal processing},
  journal       = {PRX Quantum},
  volume        = {5},
  pages         = {020368},
  year          = {2024},
  doi           = {10.1103/PRXQuantum.5.020368}
}

@misc{Sunderhauf2023,
  author        = {S\"underhauf, C.},
  title         = {Generalized quantum singular value transformation},
  year          = {2023},
  eprint        = {2312.00723},
  archivePrefix = {arXiv},
  primaryClass  = {quant-ph}
}

@article{BerntsonSunderhauf2025,
  author        = {Berntson, B. K. and S\"underhauf, C.},
  title         = {Complementary polynomials in quantum signal processing},
  journal       = {Commun. Math. Phys.},
  volume        = {406},
  pages         = {161},
  year          = {2025},
  doi           = {10.1007/s00220-025-05302-9}
}

@article{WocjanAbeyesinghe2008,
  author        = {Wocjan, P. and Abeyesinghe, A.},
  title         = {Speedup via quantum sampling},
  journal       = {Phys. Rev. A},
  volume        = {78},
  pages         = {042336},
  year          = {2008},
  doi           = {10.1103/PhysRevA.78.042336}
}

@article{RossSelinger2016,
  author        = {Ross, N. J. and Selinger, P.},
  title         = {Optimal ancilla-free {Clifford}+{$T$} approximation of $z$-rotations},
  journal       = {Quantum Inf. Comput.},
  volume        = {16},
  pages         = {901--953},
  year          = {2016},
  doi           = {10.26421/QIC16.11-12-1}
}

@article{YoderLowChuang2014,
  author        = {Yoder, T. J. and Low, G. H. and Chuang, I. L.},
  title         = {Fixed-point quantum search with an optimal number of queries},
  journal       = {Phys. Rev. Lett.},
  volume        = {113},
  pages         = {210501},
  year          = {2014},
  doi           = {10.1103/PhysRevLett.113.210501}
}

@inproceedings{Szegedy2004,
  author        = {Szegedy, M.},
  title         = {Quantum speed-up of {Markov} chain based algorithms},
  booktitle     = {45th Annual IEEE Symposium on Foundations of Computer Science},
  publisher     = {IEEE},
  address       = {Piscataway, NJ},
  pages         = {32--41},
  year          = {2004},
  doi           = {10.1109/FOCS.2004.53}
}

@article{Montanaro2015,
  author        = {Montanaro, A.},
  title         = {Quantum speedup of {Monte Carlo} methods},
  journal       = {Proc. R. Soc. A},
  volume        = {471},
  pages         = {20150301},
  year          = {2015},
  doi           = {10.1098/rspa.2015.0301}
}

@article{Arunachalam2022,
  author        = {Arunachalam, S. and Havlicek, V. and Nannicini, G. and Temme, K. and Wocjan, P.},
  title         = {Simpler (classical) and faster (quantum) algorithms for {Gibbs} partition functions},
  journal       = {Quantum},
  volume        = {6},
  pages         = {789},
  year          = {2022},
  doi           = {10.22331/q-2022-09-01-789}
}

@article{Layden2023,
  author        = {Layden, D. and Mazzola, G. and Mishmash, R. V. and Motta, M. and Wocjan, P. and Kim, J.-S. and Sheldon, S.},
  title         = {Quantum-enhanced {Markov} chain {Monte Carlo}},
  journal       = {Nature (London)},
  volume        = {619},
  pages         = {282--287},
  year          = {2023},
  doi           = {10.1038/s41586-023-06095-4}
}

@article{Lemieux2020,
  author        = {Lemieux, J. and Heim, B. and Poulin, D. and Svore, K. and Troyer, M.},
  title         = {Efficient quantum walk circuits for {Metropolis}-{Hastings} algorithm},
  journal       = {Quantum},
  volume        = {4},
  pages         = {287},
  year          = {2020},
  doi           = {10.22331/q-2020-06-29-287}
}

@article{ChenKastoryanoNature2025,
  author        = {Chen, C.-F. and Kastoryano, M. and Brand{\~a}o, F. G. S. L. and Gily\'en, A.},
  title         = {Efficient quantum thermal simulation},
  journal       = {Nature (London)},
  volume        = {646},
  pages         = {561--566},
  year          = {2025},
  doi           = {10.1038/s41586-025-09583-x}
}

@article{Ding2025,
  author        = {Ding, Z. and Li, B. and Lin, L.},
  title         = {Efficient quantum {Gibbs} samplers with {Kubo}--{Martin}--{Schwinger} detailed balance condition},
  journal       = {Commun. Math. Phys.},
  volume        = {406},
  pages         = {67},
  year          = {2025},
  doi           = {10.1007/s00220-025-05235-3}
}

@article{EberleLorler2026,
  author        = {Eberle, A. and L\"orler, F.},
  title         = {Non-reversible lifts of reversible diffusion processes and relaxation times},
  journal       = {Probab. Theory Relat. Fields},
  volume        = {194},
  pages         = {173--203},
  year          = {2026},
  doi           = {10.1007/s00440-024-01308-x}
}

@article{Grover2005,
  author        = {Grover, L. K.},
  title         = {Fixed-point quantum search},
  journal       = {Phys. Rev. Lett.},
  volume        = {95},
  pages         = {150501},
  year          = {2005},
  doi           = {10.1103/PhysRevLett.95.150501}
}

@inproceedings{Grover1996,
  author        = {Grover, L. K.},
  title         = {A fast quantum mechanical algorithm for database search},
  booktitle     = {Proceedings of the 28th Annual ACM Symposium on Theory of Computing},
  publisher     = {ACM},
  address       = {New York},
  pages         = {212--219},
  year          = {1996},
  doi           = {10.1145/237814.237866}
}

@incollection{Brassard2002,
  author        = {Brassard, G. and H{\o}yer, P. and Mosca, M. and Tapp, A.},
  title         = {Quantum amplitude amplification and estimation},
  booktitle     = {Quantum Computation and Information},
  series        = {Contemporary Mathematics},
  volume        = {305},
  publisher     = {American Mathematical Society},
  address       = {Providence, RI},
  pages         = {53--74},
  year          = {2002},
  doi           = {10.1090/conm/305/05215}
}

@article{WillsSchon2023,
  author        = {Wills, A. G. and Sch\"on, T. B.},
  title         = {Sequential {Monte Carlo}: A unified review},
  journal       = {Annu. Rev. Control Robot. Auton. Syst.},
  volume        = {6},
  pages         = {159--182},
  year          = {2023},
  doi           = {10.1146/annurev-control-042920-015119}
}

@article{Neal2001,
  author        = {Neal, R. M.},
  title         = {Annealed importance sampling},
  journal       = {Stat. Comput.},
  volume        = {11},
  pages         = {125--139},
  year          = {2001},
  doi           = {10.1023/A:1008923215028}
}

@inproceedings{Ozols2012,
  author        = {Ozols, M. and Roetteler, M. and Roland, J.},
  title         = {Quantum rejection sampling},
  booktitle     = {Proceedings of the 3rd Innovations in Theoretical Computer Science Conference},
  pages         = {290--308},
  year          = {2012},
  doi           = {10.1145/2090236.2090261}
}

@article{Hastings1970,
  author        = {Hastings, W. K.},
  title         = {{Monte Carlo} sampling methods using {Markov} chains and their applications},
  journal       = {Biometrika},
  volume        = {57},
  pages         = {97--109},
  year          = {1970},
  doi           = {10.1093/biomet/57.1.97}
}

@misc{ChowdhurySubasiSomma2018,
  author        = {Chowdhury, A. N. and Suba\c{s}\i, Y. and Somma, R. D.},
  title         = {Improved implementation of reflection operators},
  year          = {2018},
  eprint        = {1803.02466},
  archivePrefix = {arXiv}
}

@misc{ClaudonReflect2025,
  author        = {Claudon, B.},
  title         = {A simple algorithm to reflect through eigenspaces of unitaries},
  year          = {2025},
  eprint        = {2412.09320},
  archivePrefix = {arXiv}
}

@book{LevinPeres2017,
  author        = {Levin, D. A. and Peres, Y.},
  title         = {Markov Chains and Mixing Times},
  edition       = {2},
  publisher     = {American Mathematical Society},
  address       = {Providence, RI},
  year          = {2017},
  doi           = {10.1090/mbk/107}
}

@inproceedings{Gao2020,
  author        = {Gao, X. and G\"urb\"uzbalaban, M. and Zhu, L.},
  title         = {Breaking reversibility accelerates {Langevin} dynamics for non-convex optimization},
  booktitle     = {Advances in Neural Information Processing Systems},
  volume        = {33},
  year          = {2020},
  url           = {https://proceedings.neurips.cc/paper/2020/hash/cebd648f9146a6345d604ab093b02c73-Abstract.html}
}

@article{Gouraud2025,
  author        = {Gouraud, N. and Le Bris, P. and Majka, A. and Monmarch\'e, P.},
  title         = {{HMC} and underdamped {Langevin} united in the unadjusted convex smooth case},
  journal       = {SIAM/ASA J. Uncertain. Quantif.},
  volume        = {13},
  pages         = {278--303},
  year          = {2025},
  doi           = {10.1137/23M1608963}
}

@article{ClaudonRodenas2026,
  author        = {Claudon, B. and Rodenas-Ruiz, P. and Piquemal, J.-P. and Monmarch\'e, P.},
  title         = {Quantum circuits for the {Metropolis}--{Hastings} algorithm},
  journal       = {J. Phys. A: Math. Theor.},
  volume        = {59},
  pages         = {305304},
  year          = {2026},
  doi           = {10.1088/1751-8121/ae8983}
}

@article{Martyn2021,
  author        = {Martyn, J. M. and Rossi, Z. M. and Tan, A. K. and Chuang, I. L.},
  title         = {Grand unification of quantum algorithms},
  journal       = {PRX Quantum},
  volume        = {2},
  pages         = {040203},
  year          = {2021},
  doi           = {10.1103/PRXQuantum.2.040203}
}

@article{ChiangNagajWocjan2010,
  author        = {Chiang, C.-F. and Nagaj, D. and Wocjan, P.},
  title         = {Efficient circuits for quantum walks},
  journal       = {Quantum Inf. Comput.},
  volume        = {10},
  pages         = {420--434},
  year          = {2010},
  doi           = {10.26421/QIC10.5-6-4}
}

@article{LokeWang2017,
  author        = {Loke, T. and Wang, J. B.},
  title         = {Efficient quantum circuits for {Szegedy} quantum walks},
  journal       = {Ann. Phys. (N.Y.)},
  volume        = {382},
  pages         = {64--84},
  year          = {2017},
  doi           = {10.1016/j.aop.2017.04.006}
}

@article{Camps2024,
  author        = {Camps, D. and Lin, L. and Van Beeumen, R. and Yang, C.},
  title         = {Explicit quantum circuits for block encodings of certain sparse matrices},
  journal       = {SIAM J. Matrix Anal. Appl.},
  volume        = {45},
  pages         = {801--827},
  year          = {2024},
  doi           = {10.1137/22M1484298}
}

@article{EremenkoYuditskii2007,
  author        = {Eremenko, A. and Yuditskii, P.},
  title         = {Uniform approximation of $\operatorname{sgn}(x)$ by polynomials and entire functions},
  journal       = {J. Anal. Math.},
  volume        = {101},
  pages         = {313--324},
  year          = {2007},
  doi           = {10.1007/s11854-007-0011-3}
}

@article{StefankovicVempalaVigoda2009,
  author        = {\v{S}tefankovi\v{c}, D. and Vempala, S. and Vigoda, E.},
  title         = {Adaptive simulated annealing: A near-optimal connection between sampling and counting},
  journal       = {J. ACM},
  volume        = {56},
  pages         = {18},
  year          = {2009},
  doi           = {10.1145/1516512.1516520}
}

@article{WocjanChiangNagajAbeyesinghe2009,
  author        = {Wocjan, P. and Chiang, C.-F. and Nagaj, D. and Abeyesinghe, A.},
  title         = {Quantum algorithm for approximating partition functions},
  journal       = {Phys. Rev. A},
  volume        = {80},
  pages         = {022340},
  year          = {2009},
  doi           = {10.1103/PhysRevA.80.022340}
}

@inproceedings{HarrowWei2020,
  author        = {Harrow, A. W. and Wei, A. Y.},
  title         = {Adaptive quantum simulated annealing for {Bayesian} inference and estimating partition functions},
  booktitle     = {Proceedings of the 2020 ACM--SIAM Symposium on Discrete Algorithms},
  publisher     = {SIAM},
  address       = {Philadelphia, PA},
  pages         = {193--212},
  year          = {2020},
  doi           = {10.1137/1.9781611975994.12}
}

@misc{ChenKastoryanoGilyenExact2025,
  author        = {Chen, C.-F. and Kastoryano, M. J. and Gily\'en, A.},
  title         = {An efficient and exact noncommutative quantum {Gibbs} sampler},
  year          = {2025},
  eprint        = {2311.09207},
  archivePrefix = {arXiv}
}

@inproceedings{AharonovTaShma2003,
  author        = {Aharonov, D. and Ta-Shma, A.},
  title         = {Adiabatic quantum state generation and statistical zero knowledge},
  booktitle     = {Proceedings of the 35th ACM Symposium on Theory of Computing},
  publisher     = {ACM},
  address       = {New York},
  pages         = {20--29},
  year          = {2003},
  doi           = {10.1145/780542.780546}
}

@inproceedings{Gilyen2019,
  author        = {Gily{\'e}n, Andr{\'a}s and Su, Yuan and Low, Guang Hao and Wiebe, Nathan},
  title         = {Quantum singular value transformation and beyond: Exponential improvements for quantum matrix arithmetics},
  booktitle     = {Proceedings of the 51st Annual ACM SIGACT Symposium on Theory of Computing},
  publisher     = {ACM},
  address       = {New York},
  pages         = {193--204},
  year          = {2019},
  doi           = {10.1145/3313276.3316366}
}

@article{Hoeffding1963,
  author = {Hoeffding, W.},
  title = {Probability inequalities for sums of bounded random variables},
  journal = {J. Am. Stat. Assoc.},
  volume = {58},
  pages = {13--30},
  year = {1963},
  doi = {10.1080/01621459.1963.10500830}
}

@article{DavisKahan1970,
  author  = {Davis, C. and Kahan, W. M.},
  title   = {The Rotation of Eigenvectors by a
             Perturbation. III},
  journal = {SIAM J. Numer. Anal.},
  volume  = {7},
  pages   = {1--46},
  year    = {1970},
  doi     = {10.1137/0707001}
}
